\documentclass[preprint,12pt,3p,authoryear,natbib=True]{elsarticle}

\usepackage{booktabs}
\usepackage{makecell}
\usepackage{url}
\usepackage{graphicx}
\usepackage{balance}
\usepackage{amsmath}
\usepackage{subfigure}
\usepackage{graphicx}
\usepackage{balance}
\usepackage{color}
\usepackage{comment}
\usepackage{array}
\usepackage{lipsum}
\usepackage{caption}
\usepackage{pdfpages}
\usepackage{lineno}

\usepackage{blindtext}
\usepackage{enumitem}
\usepackage{lipsum}
\usepackage{capt-of}
\usepackage{tikz}
\usetikzlibrary{arrows,positioning,automata}
\usepackage{balance}
\usepackage{amsmath}
\usepackage[T1]{fontenc}
\usepackage{algpseudocode}
\usepackage{algorithm}
\usepackage{subfigure}
\usepackage{graphicx}
\usepackage[utf8]{inputenc}
\usepackage{balance}
\usepackage{multirow}
\usepackage{color}
\usepackage{comment}
\usepackage{hyperref}
\usepackage{array}
\usepackage{soul}
\usepackage[justification=centering]{caption}
\usepackage[lighttt]{lmodern}
\usepackage{libertine}
\usepackage{amssymb}
\usepackage{float}

\journal{Computers and Electronics in Agriculture}

\begin{document}
\begin{frontmatter}


\title{KisanQRS: A Deep Learning-based Automated Query-Response System for Agricultural Decision-Making}

\author[label1]{Mohammad Zia Ur Rehman\corref{cor1}}
\address[label1]{Indian Institute of Technology Indore, India}


\ead{phd2101201005@iiti.ac.in}


\author[label2]{Devraj Raghuvanshi}
\address[label2]{Shri Govindram Seksaria Institute of Technology and Science, Indore, India}
\ead{gs2019042@sgsitsindore.in}

\author[label1]{Nagendra Kumar}
\ead{nagendra@iiti.ac.in}
\cortext[cor1]{Corresponding author}


\begin{abstract}
Delivering prompt information and guidance to farmers is critical in agricultural decision-making. Farmers’ helpline centres are heavily reliant on the expertise and availability of call centre agents, leading to inconsistent quality and delayed responses. To this end, this article presents Kisan Query Response System (KisanQRS), a Deep Learning-based robust query-response framework for the agriculture sector. KisanQRS integrates semantic and lexical similarities of farmers’ queries and employs a rapid threshold-based clustering method. The clustering algorithm is based on a linear search technique to iterate through all queries and organize them into clusters according to their similarity. For query mapping, LSTM is found to be the optimal method. Our proposed answer retrieval method clusters candidate answers for a crop, ranks these answer clusters based on the number of answers in a cluster, and selects the leader of each cluster. The dataset used in our analysis consists of a subset of 34 million call logs from the Kisan Call Centre (KCC), operated under the Government of India. We evaluated the performance of the query mapping module on the data of five major states of India with 3,00,000 samples and the quantifiable outcomes demonstrate that KisanQRS significantly outperforms traditional techniques by achieving 96.58\% top F1-score for a state. The answer retrieval module is evaluated on 10,000 samples and it achieves a competitive NDCG score of 96.20\%. KisanQRS is useful in enabling farmers to make informed decisions about their farming practices by providing quick and pertinent responses to their queries.

\end{abstract}


\end{frontmatter}

\section{Introduction}
\label{sec:sa_int}
Developing nations are working towards utilizing advancements in technology to gain insights into the challenges faced by farmers. The agricultural sector faces numerous challenges, such as unpredictable weather conditions, poor infrastructure, and pest infestation which often lead to poor agricultural yield for farmers \citep{luo2023fpga}. These challenges are compounded by the lack of timely access to information which further hinders the sector's growth. One potential strategy to overcome these challenges is to furnish timely information and guidance to farmers. Keeping this in mind, policymakers in India came up with the idea of Kisan Call Centres (KCCs). These farmers' helpline centres were established by the Government of India to provide agriculture-related information and advisory services to farmers. These centres receive thousands of queries every day from farmers seeking guidance on various issues related to farming. The KCCs have become a crucial source of information for farmers, providing them with advice on crop cultivation, pest, and disease management, market prices, and government schemes \citep{chachra2020impact,raja2022constraints}.

Although the KCCs have been successful in providing information to farmers,
the effectiveness of query resolution relies heavily on the expertise and accessibility of call centre agents. This can result in delays in responding to queries and a lack of consistency in the quality of responses. Additionally, the KCCs receive a large volume of queries, which can be overwhelming for call centre agents. An automated query-response system can tackle these challenges by providing quick and accurate responses to farmers' queries using Natural Language Processing (NLP) techniques.
The massive call logs from KCC can serve as a good knowledge base for such a query-response system. A few examples of queries and responses from KCC call logs are shown in Table-\ref{table:table_example}. Several approaches have been proposed to develop an automated system that can provide answers to farmers' queries by training on the KCC dataset\footnote{https://kcc-chakshu.icar.gov.in/}. However, none of these approaches have utilized state-of-the-art artificial intelligence-based techniques to effectively solve this problem. Furthermore, existing systems that utilize traditional techniques such as Bag of Words (BoW), and Term Frequency-Inverse Document Frequency (TF-IDF) to calculate cosine similarities between sentence vectors fail to account for the semantic meaning of sentences. A few approaches collected agriculture-related information through social media such as Twitter and Facebook to solve farming problems \citep{zipper2018agricultural}. 

\begin{table}[h!]
\centering
\captionsetup{justification=centering}
\caption{Example of queries and responses from KCC dataset}
\fontsize{10}{11}\selectfont
\begin{tabular}{clp{5cm}p{7.2cm}}
\toprule
\textbf{} & \textbf{Crop} & \textbf{QueryText} & \textbf{KccAns} \\ \midrule
1 & Garlic & How to control fungal attack in garlic & Spray to mencozeb carbendazim 35-40 grampump \\ \midrule
2 & Onion & Want to know how to increase size and production in onion crop & To increase size and production in onion crop n : p : k 0:52:34 1 kg at per acer \\ \midrule
3 & Wheat & Fungal attack in  wheat  crop & Dear farmer spray of hexaconazole 5 ec 250-400 mlacre or 20mlpump and streptocyclin 2 gram at 15 liter of water in wheat crop \\ \midrule
4& Cotton Kapas & Control of pink bollworm of cotton & To control the pink bollworm of cotton use pheromon and light \\ \midrule
5 & Chillies & Varieties of chilli  & Varieties of chilli -agni -jwala sankeshwari-32 tejswini sitara phule jyoti pant c agnirekha \\ \bottomrule
\end{tabular}
\label{table:table_example}
\end{table}

This article presents a novel query-response method, KisanQRS. Kisan is a Hindi term that stands for the farmer, and QRS is the acronym for Query-Response System. KisanQRS utilizes a time-efficient threshold-based clustering algorithm to cluster similar queries. 
A Long Short-Term Memory (LSTM) based model is used to map given queries to accurate clusters. Additionally, a rule-based approach is employed to filter and ultimately return the most appropriate top-$K$ answers. To evaluate the performance of relevant cluster label retrieval (query mapping), we conduct experiments on data from five states taken from the KCC dataset. Our proposed method outperforms traditional clustering techniques significantly. To assess the quality of answers, we conduct experiments on the ranked set of answers. KisanQRS yields promising results in providing accurate answers to farmers' queries.

The main contributions of this work are as follows:
\begin{enumerate}
\item{We propose KisanQRS,  a query response system for the agricultural domain. KisanQRS can effectively provide top-$K$ responses most relevant to the given farmer query.}

\item{KisanQRS introduces a novel and easy-to-implement query clustering approach based on semantic and  token-wise similarity, which takes significantly less time than other methods. It employs an LSTM-based Deep Learning (DL) approach to map farmers' queries to the most accurate clusters. Experimental results demonstrate the efficacy of the proposed model.}

\item{KisanQRS proposes a novel answer retrieval method. The proposed answer retrieval method is based on clustering of candidate answers for a crop, then ranking of these answer clusters, and finally electing the leader of each cluster. The proposed method achieves a high NDCG score of 96.2\%.}

\end{enumerate}

Extensive research has been conducted in the field of question-answering (QA). The existing QA methods can be broadly categorized as; 1) Knowledge-based methods and 2) Machine Learning (ML) and Deep Learning-based methods. Next, we present related works based on these two categories followed by a summary of their limitations and our research objectives formulated as problem statements.
\subsection{Knowledge-based Methods}
 These methods use structured knowledge representation frameworks, such as ontologies, to represent the relationships between concepts in a domain. Knowledge-based methods rely on domain-specific knowledge bases and reasoning mechanisms to generate answers to questions. Saxena \textit{et al.}~\citep{saxena2020improving} uses knowledge base embeddings to encode information about the entities and relations in the knowledge graph. The paper suggests that incorporating knowledge base embeddings can significantly improve the performance of multi-hop question answering over knowledge graphs. Devi \textit{et al.}~\citep{devi2017adans} presents a question-answering system that is designed specifically for the agriculture domain. This paper proposes a system that uses ontologies to represent the knowledge required to answer questions in the agriculture domain. The system is designed to handle both factoid and descriptive questions.
 Deepa \textit{et al.}~\citep{deepa2022effective} proposes a method for automatically constructing an ontology for the agriculture domain using part of speech tags and the Jaccard similarity method. Menaha \textit{et al.}~\citep{menaha2023finding} presents a new method for discovering experts on community question-answering websites by utilizing domain knowledge in combination with an exact string-matching algorithm. In addition to the aforementioned works, other studies such as \citep{anggrayni2022question,wei2022construction,das2022improvement,gaikwad2015agri} also employ knowledge-based methods for answering questions. The limitation of the works cited above is that they depend on rule-based approaches to answer questions, which restricts their ability to handle more intricate reasoning tasks. In this direction, the proposed KisanQRS takes a hybrid approach that combines DL and rule-based methods. DL method is used in query mapping, which enables KisanQRS to accurately map questions that are phrased in different ways.

\subsection{Machine Learning and Deep Learning-based Methods}
ML methods use algorithms and statistical models to learn patterns and relationships in large datasets and use this knowledge to generate answers to questions. Sarrouti \textit{et al.}~\citep{sarrouti2017machine} and Yen \textit{et al.}~\citep{yen2013support} propose a Machine Learning-based approach for question-answering that uses a variety of features such as question length, presence of specific keywords, and syntactic features to classify the question type. Zin Oo \textit{et al.}~\citep{oo2021question} proposes a two-step approach for classifying questions that involve defining the tag of each word in the question and then classifying the tags using various classification algorithms such as K-NN, Naive Bayes, Decision Tree, Random Forest, SVM, and XGBoost. The limitation of the aforementioned works is that they utilize traditional NLP techniques to extract text features for question categorization. The traditional approaches offer query features that lack contextual relevance. To this end, the KisanQRS utilizes a transformer-based encoder to generate contextual features and then combines their pairwise cosine similarity with lexical similarity to assign cluster labels to queries resulting in high-quality query clusters. 

DL techniques are extensively utilized for a diverse range of NLP tasks, including text categorization~\citep{minaee2021deep}, identification of named entities~\citep{li2020survey}, and generation of word embeddings~\citep{wu2022efficiently,wang2020survey}. Deep Learning-based question-answering (QA) methods have shown great promise in recent years and have been used to build various question-answering systems, including those used in search engines, customer support chatbots, and virtual assistants \citep{hao2022recent}. McCreery  \textit{et al.}~\citep{mccreery2020effective} perform double finetuning of BERT~\citep{devlin2019bert} for the objective of identifying medical question similarity. This approach serves as an effective intermediate task, but the substantial computational overhead of BERT renders it inappropriate for performing semantic similarity searches and unsupervised tasks such as clustering. Sakata \textit{et al.}~\citep{sakata2019faq} proposes an FAQ retrieval system that utilizes both query-question similarity and BERT-based query-answer relevance. 
For similarity computation between a user query and a question from the dataset, they use the TSUBAKI~\citep{shinzato2012tsubaki} method, which relies more heavily on the syntactic structure of the text, which is also not as effective at capturing contextual relationships between words in a query as recent transformer-based models. Arora \textit{et al.}~\citep{arora2020agribot} gives an approach for a conversational chatbot for agricultural queries. They use a sequence-to-sequence model to build the chatbot using LSTM and GRU. Bi \textit{et al.}~\citep{bi2021bi}, Chen \textit{et al.}~\citep{chen2022intelligent}, and Zhu \textit{et al.}~\citep{zhu2023question} explore the utility of attention mechanism with DL techniques for the question-answering system in different domains. Kim \textit{et al.}~\citep{kim2022question} presents a question-answering method for infrastructure damage information using Sentence BERT (SBERT)~\citep{reimers2019sentence} and BERT. SBERT is used for sentence-level information and BERT is used for token-level information in their method. Bansal \textit{et al.}~\citep{bansal2022hybrid} proposes a multimodal framework for hashtag recommendation which uses BERT for textual features with techniques like multilabel classification and sequence generation.

Other than these, a few QA methods have been suggested on  KCC dataset \citep{ajawan2020smart,mohapatra2018using, mohapatra2018query}. These approaches have certain drawbacks, including their computational intensity, as each new query needs to be compared with existing queries, rendering them impractical for handling large volumes of queries. Additionally, they rely on traditional NLP techniques such as TF-IDF and BoW which provide non-contextual features from query text.
Other works on the KCC dataset include Godara \textit{et al.} \citep{godara2022deep} which suggests the DL approach for query count prediction using the KCC dataset and Godara \textit{et al.} \citep{godara2020sequential} which gives a framework to derive pertinent association rules using Apriori algorithm.

To summarize, the aforementioned cited works have limitations in their reliance on rule-based approaches for question answering, which hampers their ability to handle complex reasoning tasks. Similarly, ML methods suffer limitations in their use of traditional techniques for extracting text features in question categorization, resulting in query features that lack contextual relevance. DL methods show improvement over rule-based and ML methods, but the cited works have used fine-tuning of BERT which has a drawback of heavy resource consumption. To this end, KisanQRS uses LSTM to take advantage of DL methods, SBERT for faster and resource-friendly inference of embeddings, and a rapid clustering approach which makes our model faster and computationally lighter as compared to existing methods.

The KCC dataset comprises data for various states and union territories. We extract a subset of the dataset that pertains to a particular state $s \in S$, where $S$ denotes the set of all states. For state $s$, we define the sets $Crops$, $Q = \{q_i\}^{N}_{i=1}$, and $A = \{a_i\}_{i=1}^N$ to represent the columns `Crop', `QueryText', and `KccAns', respectively. Here, $q_i$ denotes the $i^{th}$ query and $a_i$ represents its corresponding answer, and $N$ represents the total number of query-response pairs in the dataset. We aim to cluster the queries in the dataset into similar groups, map new queries to the most relevant clusters, and retrieve the most relevant answers based on specific ranking criteria. Hence, the following problems constitute the research objectives for this article:

\begin{enumerate}
\item \emph{Problem 1: Query Clustering}\\
Our first objective is to cluster the queries in $Q$ into semantically as well as lexically similar groups, forming the set of query clusters $C_{query} = \{\alpha_k\}^{M}_{k=1}$. Here, $\alpha_k$ represents the $k^{th}$ query cluster which is a set of similar queries, and $M$ represents the total number of query clusters formed.
It is necessary to compare different clustering techniques in order to determine which one is providing superior outcomes.

\item \emph{Problem 2: Query Mapping}\\
Our second objective is to map $q_{user}$ to the most relevant cluster $\alpha_k$ such that the queries in $\alpha_k$ are semantically similar to $q_{user}$. This problem can be approached as a supervised learning task, where the model is trained on a labeled dataset of queries and their corresponding clusters. We can preprocess the dataset to extract relevant features from each query, such as keywords, topics, etc. Next, an ML or DL model can be trained on the labeled dataset to learn the mapping between the queries and the corresponding clusters. Finally, the trained model can be used to predict the cluster for $q_{user}$ based on its feature vector.

\item \emph{Problem 3: Answer Retrieval}\\
Once $q_{user}$ has been mapped to $\alpha_k$, our third objective is to retrieve the top-$K$ answers from $\beta_k$ that are most relevant to $q_{user}$ based on a specific ranking criterion, where $\beta_k$ is the set of answers for all queries $q_i$ such that $q_i \in \alpha_k$. These retrieved answers should be for the crop mentioned in $q_{user}$, if one is specified. This objective requires us to develop a method for ranking the answers based on their relevance to the query and the crop.

\end{enumerate}

The remainder of this article is organized as follows.
Section-\ref{sec:sa_meth} provides an extensive discussion of the methodology and experimental details employed in the study. Section-\ref{sec:sa_ee} presents the results of the study. Finally, Sections-\ref{sec:sa_disc} and \ref{sec:sa_cnfw} provide a discussion and conclusion for the study, respectively. 

\section{Materials and Methods}
\label{sec:sa_meth}
This section outlines the proposed methodology and experimental details of KisanQRS. The methodology consists of four phases: Data Preprocessing, Query Clustering, Model Training, and Answer Retrieval and Ranking. Subsequently, we discuss the experimental details which include the hyperparameter and model details, evaluation metrics for different modules of KisanQRS, and the description of the dataset.

\subsection{Methodology}
\label{label:sb:meth}
\begin{figure}[h!]
    \centering
    \includegraphics[width=16cm, height=8.55cm]{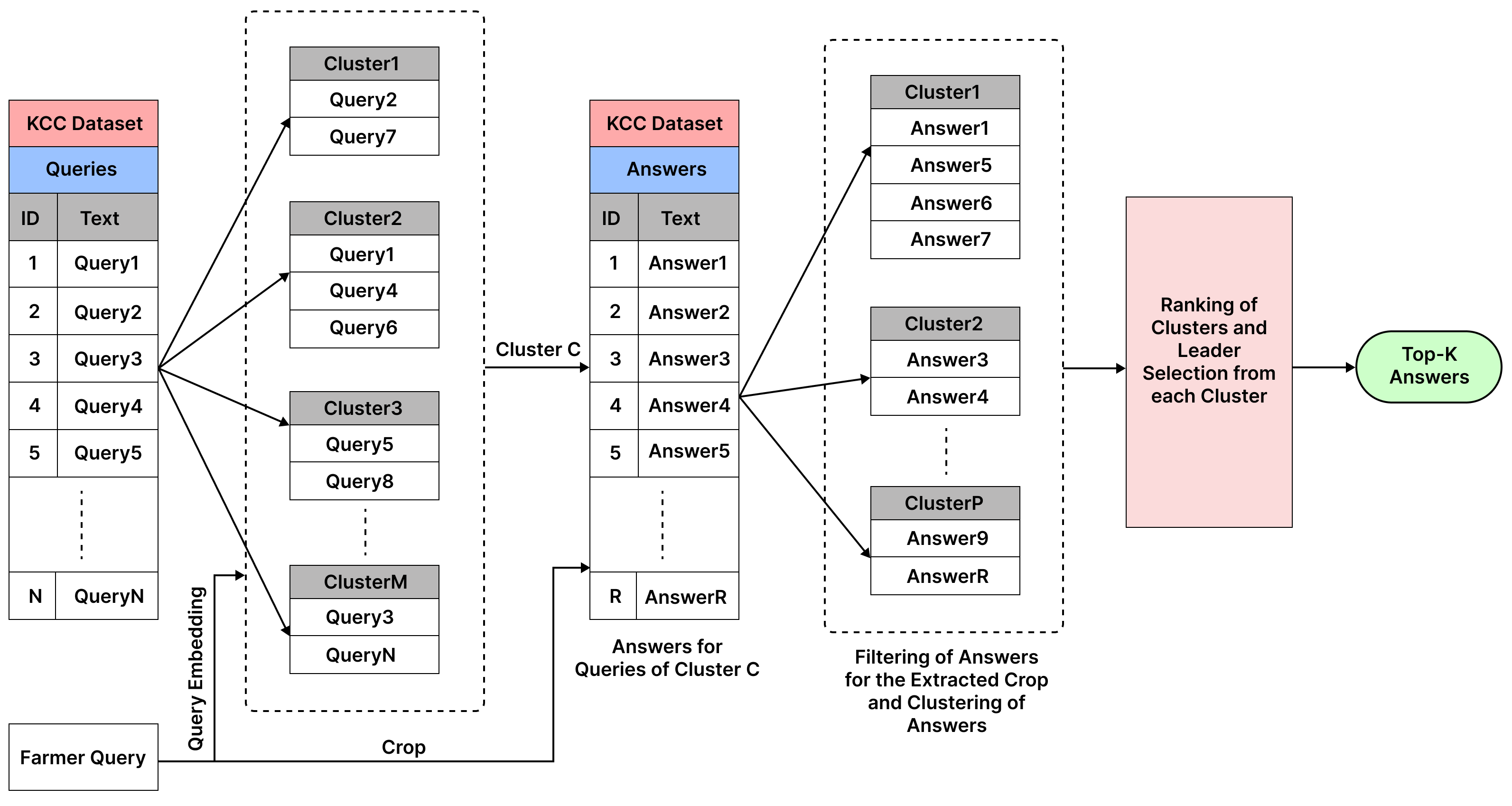}
    \caption{End-to-end Schematic Representation of KisanQRS}
    \label{fig:ansclus}
\end{figure}

A general end-to-end schematic representation of KisanQRS is shown in Figure-\ref{fig:ansclus}. First, the queries are grouped into clusters according to their semantic and token similarities by the query clustering module. Next, a query mapping module is trained to map any new queries to the most fitting cluster. Finally, potential responses are grouped into clusters and the top-$K$ most pertinent answers are selected by the answer retrieval and ranking module to serve as the conclusive answers to the farmer's query. A detailed discussion of the phases and architectures of different modules is given in the subsequent sections.

\subsubsection{Data Preprocessing}
\label{label:Prep}

This subsection discusses the data preprocessing steps that are essential for effective feature extraction. The section is divided into two main parts: data cleaning and data transformation.

\begin{enumerate}

    \item \emph{Data Cleaning:}
    \label{label:clean}
    In this step, entries that contain empty queries $q_i$ or answers $a_i$ are removed from the dataset. Then, all instances of special characters 
    including commas, hyphens, semicolons, and extra white spaces, are purged $\forall q_i \in Q$. Lastly, any duplicate rows present in the dataset are eliminated.

    \item \emph{Data Transformation:}
    \label{label:trans}
    Data transformation is the process of modifying the original data to make it more suitable for analysis or modeling. 
    Queries that ask for real-time data, such as market rates or weather conditions are removed from $Q$ as it is not possible to determine their answers based on past data. In addition, if any query $q_i \in Q$ contains a token $c \in Crops$, then $c$ is removed from $q_i$ to generate similar feature vectors for similar queries, resulting in the formation of better clusters. For example, if we have queries such as \textit{fertilizer dose for onion} and \textit{fertilizer dose for tomato}, the tokens \textit{onion} and \textit{tomato} are removed from these queries, respectively, resulting in both queries being reduced to \textit{fertilizer dose} after preprocessing. This increases the likelihood of both queries being grouped into the same cluster by the clustering technique, resulting in the formation of a cluster $\alpha_k$ containing queries for the fertilizer dose of multiple crops. Thus, when a user asks a query about the fertilizer dose for a specific crop, $q_{user}$ can be mapped to $\alpha_k$ first, and all queries related to the crop mentioned in $q_i$ can be filtered from $\alpha_k$. This approach prevents the formation of redundant clusters, which could result in lower performance of the supervised learning method due to multiple clusters addressing the same problem for different crops. We develop two versions of $Q$, $Q_{jaccard}$ and $Q_{sbert}$. In $Q_{jaccard}$, we remove stopwords and apply stemming. However, for $Q_{sbert}$, we retain the complete context as transformer-based embedding models require it to generate accurate embeddings.
    
\end{enumerate}

\subsubsection{Query Clustering}
\label{sec: query_clust}
\begin{figure}[t!]
    \centering
    \includegraphics[width=16cm, height=13.5cm]{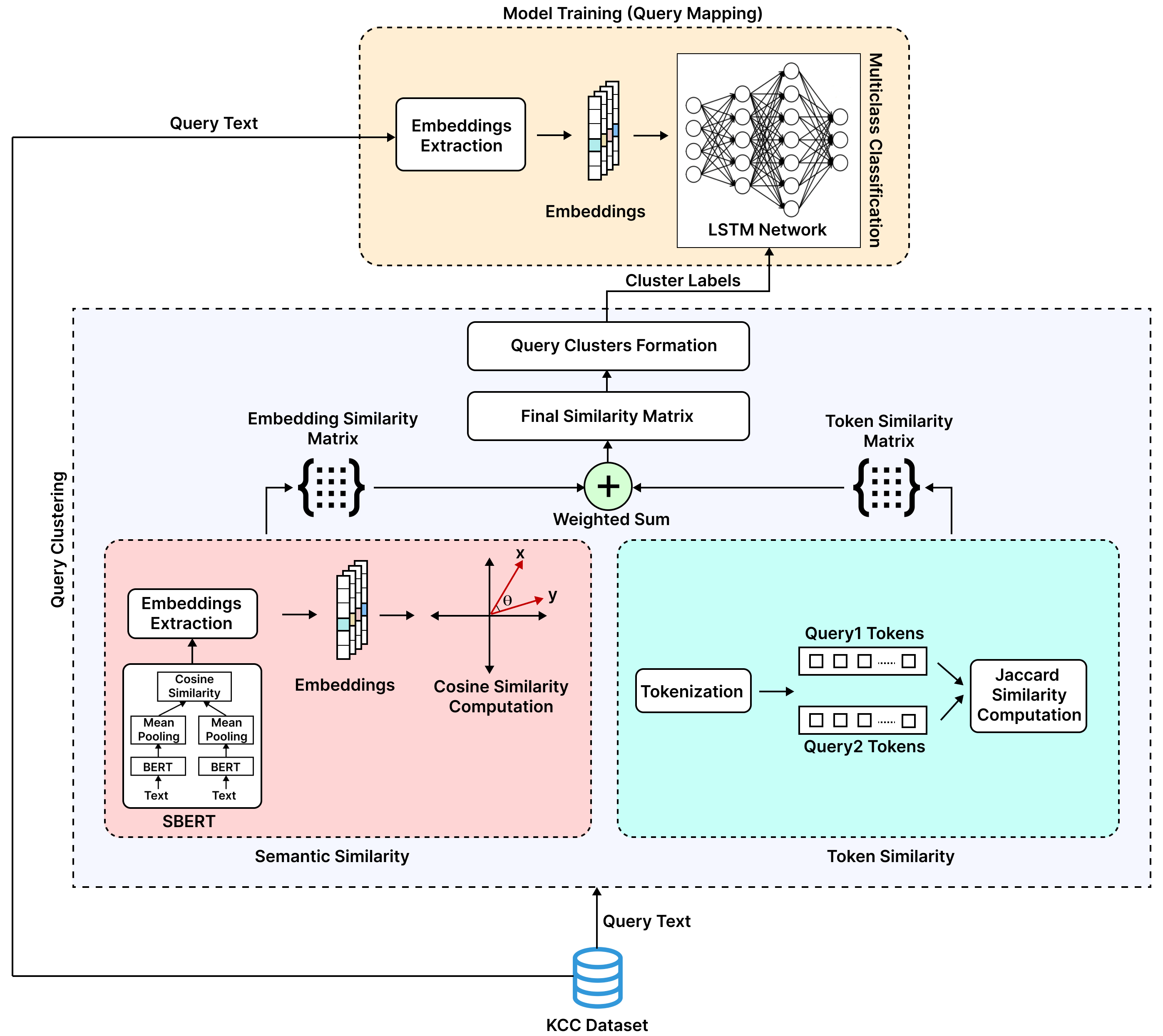}
    \caption{Query Clustering and Training Module}
    \label{fig:clustering}
\end{figure}

 Numerous entries in the dataset have semantically and lexically similar queries, so it is imperative to cluster such queries. Clusters generated in this module serve as the target labels for the DL model for query mapping.  The query clustering module has two main components: similarity matrix generation and query cluster formation as shown in Figure-\ref{fig:clustering}.

\begin{enumerate}

\item \emph{Similarity Matrix Generation:}
\label{label:sim_matrix_gen}    
The similarity between two queries is measured as the weighted sum of cosine similarity between their contextual embeddings and token-wise similarity using the Jaccard Index~\citep{jaccard1901etude}, as shown in Equation-\ref{eq:sim}.

\begin{equation}
    sim(q_i,q_j) = \lambda(sim_{emb}) + (1-\lambda)sim_{token}
\label{eq:sim}
\end{equation}

Here, $sim(q_i,q_j)$ denotes the similarity score between query $q_i$ and query $q_j$, $sim_{emb}$ denotes cosine similarity between the contextual embeddings of query $q_i$ and $q_j$, $sim_{token}$ denotes token-wise similarity and $\lambda$ denotes the weight given to embedding similarity. We conduct experiments using different values of $\lambda$, ranging from 0 to 1 with increments of 0.01. 
Empirical analysis indicates that the best results are achieved when $\lambda$ is between 0.78 and 0.81. These results are the same for any value between 0.78 to 0.81. So we select 0.8 as the final value of $\lambda$. One of the reasons for $\lambda$ yielding improved results in this range, particularly when giving higher weight to SBERT embeddings, is their stronger correlation with the cluster labels compared to the Jaccard similarity. This can be attributed to the fact that SBERT embeddings consider the context of a query, enhancing their effectiveness in capturing meaningful associations with the cluster labels. \\
To compute $sim_{emb}$, first, queries are encoded into contextual embeddings; then similarity is measured between two queries using cosine similarity. Cosine similarity is a measure of similarity between two non-zero vectors of an inner product space. Given two embeddings $emb_i$ and $emb_j$, the cosine similarity is defined as the cosine of the angle between these two vectors, and it can be computed as:

\begin{equation}
    Cosine\_similarity(emb_i,emb_j) = \frac {(emb_i\cdot emb_j)} { (\left \| emb_i \right \|)*(\left \| emb_j \right \|)}
\label{eq:cosine}
\end{equation}

where, $``\cdot"$ denotes the dot product and $\left \| x \right \|$ denotes the Euclidean norm of a vector $x$. The cosine similarity lies between -1 and 1, with 1 indicating that the vectors are perfectly similar and -1 indicating that they are completely dissimilar. Contextual embeddings are extracted using SBERT transformer encoder. The architecture of the SBERT encoder is given in Figure-\ref{fig:clustering}. For our final model, we have used SBERT for embedding extraction; hence in Figure-\ref{fig:clustering} and Figure-\ref{fig:answer}, wherever embeddings extraction is mentioned, it should be understood that SBERT is used at that location.

\textbf{SBERT}  is a variation of the pre-trained BERT model that employs siamese and triplet network structures to produce semantically rich contextual sentence embeddings. 
Siamese neural networks are unique networks composed of two or more identical sub-models that share the same parameters. The updating of the parameters is reflected in both sub-models. Although the originally proposed BERT model generates state-of-the-art embeddings, it requires a huge number of inference computations thus making it unsuitable for large-scale clustering tasks. On the other hand, SBERT derives fixed-length sentence embeddings using siamese architecture resulting in a multi-fold reduction in inference computations.

To compute $sim_{token}$, we use Jaccard similarity, which is measured as the overlap of words (tokens) between two queries. If $A$ and $B$ are the sets of tokens in two queries, the Jaccard similarity between these queries can be defined as follows:
\begin{equation}
Jaccard\_similarity(A, B) = \frac {|A \hspace{1mm} \cap \hspace{1mm} B|}{ |A \hspace{1mm} \cup \hspace{1mm} B|}
\label{eq:jaccard}
\end{equation}

The Jaccard similarity lies between 0 and 1, with 1 indicating that the queries have the exact same tokens, and 0 indicating that they have no tokens in common.\\
Above mentioned, similarities are computed for each pair of queries to generate the final similarity matrix, $sim$.

\item \emph{Query Cluster Formation:}
\label{sec: query_cluster_form}
We implement a threshold-based clustering technique to group queries into clusters. Algorithm-\ref{algo:clustering} represents an implementation of  our threshold-based clustering. This algorithm clusters data points based on their similarities, where each data point is represented as a vector of features, and the similarities between pairs of data points are represented in a similarity matrix. The input to the function $Cluster$ includes, a similarity matrix $sim$, a similarity threshold $thresh$, and a minimum size for clusters $min$. The output is a list of clusters, where each cluster is represented as a set of queries.
The algorithm uses a simple linear search approach to iterate over all queries and group them into clusters based on their similarity. The threshold value is used to determine whether two queries should be grouped together, and the minimum size ensures that only clusters with a certain number of queries are retained.

\begin{algorithm}[h!]
    \begin{tabular}{ll}
    \textit{Input:} & $sim$:  similarities matrix\\
    &  $thresh$  : similarity threshold \\
    &  $min$  : minimum size for a cluster \\
  \textit{Output:} 
  & $clusters$: list of clusters\\
  \textit{function:} & Cluster($sim, thresh, min$)
    \end{tabular}
    
\caption{Clustering}
\label{algo:clustering}
\begin{algorithmic}[1]

\Function{Cluster}{$sim$, $thresh$, $min$}
  \State $n \gets \text{len}(sim)$
  \State $visited\gets \text{zeros}(n)$
  \State $C_{query} \gets \{\}$
  \ForAll{$q_i \in Q: visited(q_i) == False$}
    \State $visited(q_i) \gets True$
    \State $\alpha_i \gets \{q_i\}$
      \ForAll{$q_j \in Q: (i+1 \leq j < n$ \& $visited(q_j) == False$ \& $sim(q_i, q_j) \ge thresh$)}
            \State $visited(q_j) \gets True$
            \State $\alpha_i.\text{add}(q_j)$
      \EndFor
      \If{$|\alpha_i| \ge min$}
        \State $C_{query}.\text{add}(\alpha_i)$
      \EndIf
  \EndFor
  \State \Return $C_{query}$
\EndFunction
\end{algorithmic}
\end{algorithm}

In Algorithm-\ref{algo:clustering}, Line 1 initializes the variable $n$ to store the length of the similarity matrix or the total number of queries. In Line 2, an array called $visited$ is initialized with zeros, where each index corresponds to the data points in the input set $Q$, and the values indicate whether or not a data point has been visited. Line 3 initializes an empty list, $C_{query}$, to store the clusters that will be found. The outer loop; Line 4-7, iterates over each data point $q_i$ in the input set $Q$. If $q_i$ has not been visited yet, we mark it as visited and create a new cluster, $\alpha_i$, with $q_i$ as its only member. The inner loop, lines 8-10, iterates over the remaining unvisited data points $q_j$ to search the data points that have a similarity greater than or equal to the threshold with $q_i$. Each unvisited data point $q_j$ that satisfies the condition is marked as visited and is added to the cluster $\alpha_i$. The inner loop terminates once all remaining unvisited data points have been considered for the current data point $q_i$. In lines 12 and 13,  if the size of the cluster $\alpha_i$ is greater than or equal to the minimum size then $\alpha_i$ is added to the list of clusters, $C_{query}$. The outer loop terminates once all data points in the input set $Q$ have been considered. The list of clusters $C_{query}$ is returned as the output of the clustering algorithm. 

In general, our proposed clustering method is related to the order of data points. Consider three queries: \textit{a}, \textit{b}, and \textit{c}. The similarity between \textit{a} and \textit{b} is greater than the threshold, and the similarity between \textit{b} and \textit{c} is also greater than the threshold. However, if the similarity between \textit{a} and \textit{c} is less than the threshold, then \textit{c} cannot be clustered with \textit{b} because \textit{a} and \textit{b} appear first and are clustered together.

However, in the KCC dataset, queries for a specific topic mostly become the same after removing crop token(s) during preprocessing. For example, queries such as \textit{yellowing on turmeric}, \textit{sugarcane yellowing}, and \textit{yellowing on melon} are reduced to only \textit{yellowing} after data preprocessing steps, and they are clustered together as shown in Table-\ref{table:cluster_examples}. Due to this characteristic of the KCC dataset, if \textit{a} and \textit{b} have a similarity higher than the threshold and \textit{b} and \textit{c} also have a similarity higher than the threshold, there is a very high probability that \textit{a} and \textit{c} will also have a similarity greater than the threshold. In other words, it is rare for the similarity between \textit{a} and \textit{c} to be less than the threshold. Therefore, we can safely ignore scenarios where this is not the case and assert that all similar queries are ultimately grouped together by our proposed algorithm despite it being related to the order of data points.

\begin{table}[h!]
\centering
\caption{Representation of Similar Keywords in Clusters}
\begin{tabular}{lll}
\toprule
\textbf{S.No.} & \textbf{QueryText} & \textbf{ClusterID} \\
\midrule
1 & \textbf{yellowing} on melon & 1 \\
2 & sugarcane \textbf{yellowing} & 1 \\
3 & \textbf{yellowing} on turmeric & 1 \\
\midrule
4 & the \textbf{nematode attack} on bean & 2 \\
5 & \textbf{attack of nematodes} on ginger & 2 \\
6 & \textbf{attack of nematode} on capsicum & 2 \\
7 & \textbf{attack of nematode} in pomegranate & 2 \\
\midrule
8 & precautions to \textbf{avoid bollworm} in cotton & 3 \\
9 & precautions to \textbf{avoid bollworm} in maize & 3 \\
10 & precautions to \textbf{avoid bollworm} in tomato & 3 \\
\bottomrule
\end{tabular}
\label{table:cluster_examples}
\end{table}

We also implement an alternate approach, where the similarity is considered with respect to each member of a cluster to determine whether a new query should be included in that cluster. However, during experiments, we observe that the alternate algorithm's runtime is significantly higher than our proposed algorithm while providing minimal improvement in results. Hence, we conclude that our proposed algorithm is better suited for this dataset.

Overall, our proposed algorithm provides a framework for clustering queries based on semantic and lexical similarities. It is an effective algorithm that is fast and simple to implement. The proposed clustering algorithm significantly outperforms K-Means, Agglomerative, and DBSCAN clustering algorithms, as shown in Table-\ref{table:table_cluster}.

\textit{Time Complexity of Clustering :}\\
The time complexity of the given algorithm is $O(n^2)$, where n is the number of data points.
The outer loop (lines 5-15) iterates over all data points, resulting in $O(n)$ iterations. The inner loop (lines 8-11) also iterates over all remaining data points, resulting in an average of less than $O(n/2)$ iterations per outer loop iteration. Since the two loops are nested, approximately the total number of iterations is $O(n * n/2)$, which simplifies to $O(n^2)$.

\subsubsection{Model Training}
\label{sec: training}

A DL model is trained with the supervised multi-class classification objective for query mapping. Query mapping is the process of assigning an accurate cluster label to a farmer's query. In this task, the query embeddings extracted using a transformer encoder serve as the input, while the cluster labels act as the target. 
We evaluate multiple models to identify the most suitable architecture for our task. Among the different options considered, the LSTM model consistently demonstrates superior performance and exhibits the ability to effectively capture temporal dependencies in the data. As a result, we select LSTM for the query mapping module. The detailed comparison and performance analysis of other models explored during the experimentation can be found in Section-\ref{subsubsec:model_selection}. \\

\begin{table}[h!]
\centering
\caption{Model Architecture and Hyperparameter Details}
\resizebox{\textwidth}{!}{\begin{tabular}{l l l l l l}
\toprule
\multicolumn{3}{c}{\textbf{Model Architecture}} & \multicolumn{3}{c}{\textbf{Hyperparameters}} \\ \cmidrule(lr){1-3} \cmidrule(lr){4-6}
\textbf{Layer} & \textbf{Activation Function} & \textbf{Output Dimension} & \textbf{Hyperparameter} & \textbf{Value} &  \\ \cmidrule(lr){1-3} \cmidrule(lr){4-6}
LSTM  & tanh                & 768              & Learning rate  & $(\mathbf{10^{-3}}, 10^{-4}, 10^{-5})$ &  \\
LSTM  & tanh                & 512              & Batch size     & $(16, \mathbf{32}, 64, 128)$   &   \\
Dense & softmax             & Number of clusters   & Epochs         & $(10, 20, \mathbf{30}, 50, 100)$ & \\
       &                     &                  & Dropout        & $(0.1, \mathbf{0.2}, 0.3, 0.4, 0.5)$ & \\ &  &  &  Optimizer        & Adam & \\
\bottomrule
\end{tabular}}
\label{table:experimental_setup}
\end{table}

Table-\ref{table:experimental_setup} represents the model architecture and hyperparameter details utilized in our experimental setup. The \textit{Model Architecture} section highlights the model's layers, activation functions, and output dimensions. The input features extracted using SBERT have a dimensionality of 768. Considering this, we set the output dimension of the first LSTM layer to be 768 to preserve the initial information and allow for effective information processing within the LSTM. For the next LSTM layer, an output dimension of 512 is chosen to capture the essential information while avoiding potential overfitting or excessive computational requirements. The \textit{Hyperparameter} section outlines the specific values experimented with for learning rate, batch size, epochs, dropout, and optimizer. By systematically exploring various hyperparameter settings, our aim was to identify the optimal configuration and enhance model training and performance. The values of hyperparameters that yielded the best results are emphasized in bold. These values represent the final hyperparameter configuration adopted in our model.

\subsubsection{Answer Retrieval and Ranking}
This task is divided into two submodules: query  mapping module, and answer retrieval module as shown in Figure-\ref{fig:answer}.

\begin{figure}[h!]
    \centering
    \includegraphics[width=16cm, height=4cm]{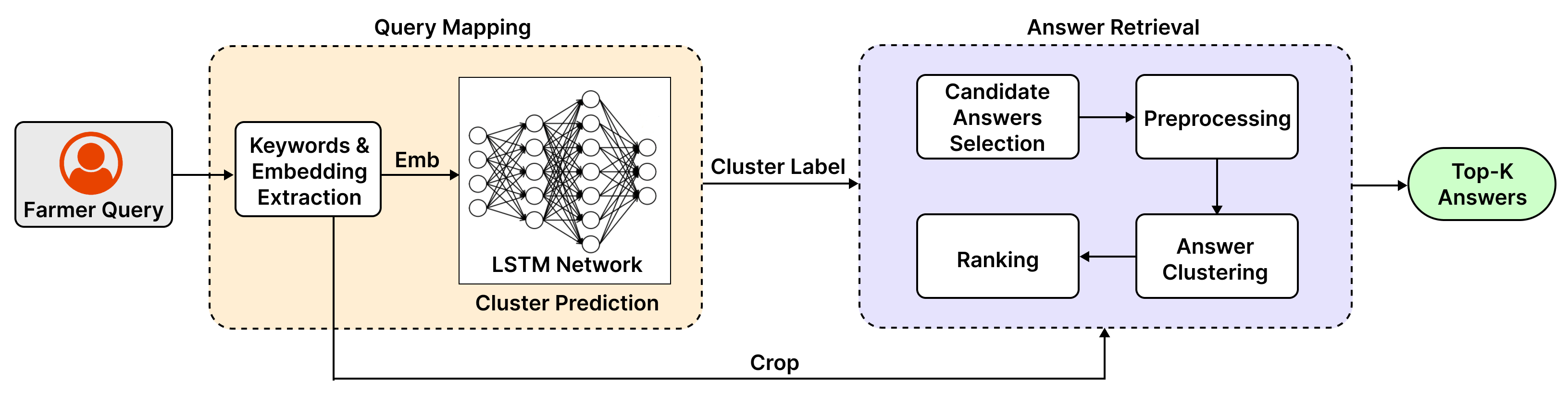}
    \caption{Answer Retrieval and Ranking Module}
    \label{fig:answer}
\end{figure}

\begin{enumerate}

\item \emph{Query Mapping:}
The user query $q_{user}$ is first preprocessed, then the contextual features are extracted from $q_{user}$ which serve as an input to the LSTM model for inference.
\begin{enumerate}
    \item \emph{Preprocessing User Query}:
    In this phase, special characters are removed from $q$ and crop detection is performed, in which let's say a crop $c \in Crops$ is detected i.e. $\exists$ $c \in $ N-grams$(q_{user}): c \in Crops$, which is also removed from $q_{user}$ due to the reason explained in Section-\ref{label:Prep}.
    \item \emph{Feature Extraction and Inference}:
    From the preprocessed query, the contextual features are extracted using the SBERT encoder. The features are then served as an input to the LSTM model for inference.
    Let's say that $q_{user}$ is mapped to cluster $\alpha_k = \{q_0^k, q_1^k, q_2^k, ..., q_{n-1}^k\}$ by the model. 
    The query clusters are formed such that each query $q_x^k$ has a similar meaning as well as a lexical structure similar to the rest of the queries in $\alpha_k$, which means if $q_{user}$ has been mapped to $\alpha_k$, then each answer $a_x^k$ will be a correct answer for $q_{user}$ such that crop($a_x^k$) $= c$.\\
    Here, $q_x^k$ denotes a query of cluster $\alpha_k$, where $x$ is used to index the queries or their corresponding answers and $n$ is the total number of queries in $\alpha_k$. For each query $q_x^k$, the corresponding answer is denoted as $a_x^k$ and $\beta_k$ denotes the set $\{a_0^k, a_1^k, a_2^k, ..., a_{n-1}^k\}$.
\end{enumerate}

\item \emph{Answer Retrieval:}
Given a user query $q_{user}$ and the cluster to which it is mapped ($\alpha_k$), the next task is to retrieve top-$K$ answers from the set $\beta_k$. This task is divided into three phases which are elaborated below:
\begin{enumerate}
    \item \emph{Candidate Answer Selection}:
    Once the query has been mapped to the cluster $\alpha_k$, some answers $a_x^k$ for the queries in $\alpha_k$ might not be for the crop $c$ that the user is querying for. Out of all these candidate answers $A$ in the predicted cluster $\alpha_k$, only the candidate answers containing the crop $c$ (which is detected during the preprocessing stage) are selected which form the set $\beta_k^{'}$.\\
    If the user has not mentioned any crop in the query, then the filtering on the basis of the crop is not performed and all the answers in $\beta_k$ are taken into account, i.e. $\beta_k^{'} = \beta_k$ for this case.

 \begin{algorithm}
        \begin{tabular}{ll}
    \textit{Input:} & $ques$:  user query\\
    &  $K$  : number of answers \\
  \textit{Output:} 
  & $leaders$: top-$K$ answers\\
  \textit{function:} & top\_K\_ans($ques, K$)
    \end{tabular}

\caption{Answer Retrieval and Ranking}
\label{algorithm:top_K_ans}
\begin{algorithmic}[1]
\Function{top\_k\_ans}{$ques, K$}
 
    \State $leaders$ $\gets$ []
    \State $\beta_k$ $\gets$ answers in the predicted question cluster
     \State $\beta_k^{'}$ $\gets$ \{$a_{\lambda}:a_{\lambda} \in \beta_k$ and $\exists$ token in $a_{\lambda}: $ token $\in Crops$\}
     \State $sim$ $\gets$ $\{s_{xy}|s_{xy} = \frac{sim_{char}(x, y) + sim_{token}(x, y)}{2}; x,y \in \beta_k^{'}\}$
    \State $C_k$ $\gets$ Cluster($sim$, $thresh$, $min$)
    \State $C_k^{'}$ $\gets$ $C_k$ sorted in decreasing order (criteria: cluster size)
    \ForAll{$\gamma_i \in C_k^{'}$}
    \State $L_i$ $\gets$ leader($\gamma_i$)
    \State $leaders$.append($L_i$)
    \If{len($leaders$) == $K$}\ break \EndIf
    \EndFor
  \State \Return $leaders$
\EndFunction
\end{algorithmic}
\end{algorithm}

    \item \emph{Similarity Matrix Generation and Clustering}:
    Within the set $\beta_k^{'}$, certain answers exhibit strong semantic similarity with each other. So, in order to extract a single answer from these groups of similar answers, we first perform clustering $\forall a_x^k \in \beta_k^{'}$ to obtain a set of answer clusters $C_k = \{\gamma_x\}_{x=1}^P$, where $P$ denotes the total number of answer clusters formed. For making answer clusters, we first generate the similarity matrix for all the answers $a_x^k \in \beta_k^{'}$. The similarity between two answers $a_1$ and $a_2$ is measured as the average of character similarity and token-wise similarity between $a_1$ and $a_2$ as shown in Equation-\ref{eq:ans_sim}.

    \begin{equation}
    sim(a_1,a_2) = \frac{sim_{char}(a_1, a_2) + sim_{token}(a_1, a_2)}{2}
    \label{eq:ans_sim}
    \end{equation}

    Here, $sim_{char}$ denotes the character-wise similarity between $a_1$ and $a_2$ and $sim_{token}$ denotes the token-wise similarity between $a_1$ and $a_2$ and both are calculated using Jaccard index.

    After the similarity matrix generation, clustering is performed $\forall a_x^k \in \beta_k^{'}$, similarly as performed in Section-\ref{sec: query_clust} using Algorithm-\ref{algo:clustering}.

    \item \emph{Ranking top-$K$ Answers}:
    To select the top-$K$ answers from $\beta_k^{'}$, we first rank the answer clusters where
    the ranking of answer cluster $\gamma_i$ is determined by its size $|\gamma_i|$ in comparison to other answer clusters. The larger the size, the higher the rank of $\gamma_i$. Subsequently, from each answer cluster, a leader answer is chosen. As the answers in $\gamma_i$ are expected to be similar due to being part of the same cluster, the leader $L_i$ of $\gamma_i$ is chosen based on the answer with the most keywords coverage. The leader conveys the most comprehensive information among all the answers in $\gamma_i$. Finally, the top-$K$ leaders are returned based on the ranking of answer clusters.
    Algorithm-\ref{algorithm:top_K_ans} gives steps for answer retrieval and ranking.
\end{enumerate}
      
\end{enumerate}

\subsection{Experimental Details}
In this section, we discuss the evaluation metrics utilized for assessing the performance of KisanQRS, along with comprehensive information regarding the dataset employed in our experiments.

\subsubsection{Evaluation Metrics}
In this section, we discuss the evaluation metrics for different modules of KisanQRS: Query Clustering, Query Mapping, and Answer Retrieval and Ranking.
\begin{enumerate}
\item{\textit{Evaluation Metrics for Clustering Method}}\\
We use Silhouette Score, Calinski-Harabasz (CH) Index, and Davies-Bouldin (DB) Index to evaluate the performance of our clustering method. Their brief description is given as follows:
\begin{enumerate}
    \item{\textit{Silhouette Score }: This metric is utilized in cluster analysis for assessing the quality of the formed clusters. Consider $a_i$ as the mean distance between the $i^{th}$ data point and all other data points that belong to the same cluster. Similarly, let $b_i$ refer to the mean distance between the $i^{th}$ data point and all the data points present in the closest cluster (i.e., the cluster with the minimum average distance to the given data point), then Silhouette score, $s_i$, for a single data point is as follows:
    
\begin{equation}
    s_i = \frac{b_i - a_i}{\max{a_i, b_i}}
\end{equation}
    If the Silhouette score is high, it implies that a data point fits well in its assigned cluster but not in neighboring clusters. Conversely, a low score indicates the opposite, that is, a data point may be a better fit for a neighboring cluster than its assigned cluster. The overall Silhouette Score for the clustering solution is calculated as the average Silhouette score of all data points in all clusters.}

\item{\textit{Calinski-Harabasz Index }: It measures the ratio of between-cluster dispersion and within-cluster dispersion, where higher scores indicate better clustering results. The mathematical formula for CH Index is:

\begin{equation}
    CH(n) = \frac{Tr(B)}{Tr(W)} \times \frac{N - n}{n - 1}
\end{equation}

Here, $n$ represents the number of clusters, $N$ represents the total number of data points, $W$ represents the within-cluster dispersion matrix (also known as the sum of squared errors within each cluster), and $B$ represents the between-cluster dispersion matrix (also known as the sum of squared errors between each cluster). $Tr$ denotes the trace of a matrix, which is the sum of its diagonal elements.}
\item{\textit{Davies-Bouldin Index }: It is used to evaluate clustering quality by calculating the average similarity between each cluster and the cluster that is most similar to it. Its objective is to minimize the intra-cluster distance while maximizing the inter-cluster distance. The mathematical formula for the Davies-Bouldin index can be expressed as:

\begin{equation}
    DB = \frac{1}{n}\sum_{i=1}^{n}\max_{j \neq i}(\frac{s_i + s_j}{d(c_i,c_j)})
\end{equation}

Here, $n$ refers to the number of clusters, $c_i$ represents the centroid of the $i^{th}$ cluster, $s_i$ denotes the average distance between each point in the $i^{th}$ cluster and its centroid, and $d(c_i, c_j)$ is the distance between the centroids of the $i^{th}$ and $j^{th}$ clusters. The Davies-Bouldin index ranges from 0 to infinity, where a score of 0 indicates perfect clustering, while a higher score indicates poorer clustering.}
\end{enumerate}

\item{\textit{Evaluation Metrics for Query Mapping Module}}\\
 Two of sets of investigations are conducted for the query mapping module. 
 A detailed discussion of these experiments is presented in the next section, i.e. Section-\ref{subsec:query_mapping results}. In this section, we discuss the different evaluation metrics used for the query mapping module. Precision, Recall, F1-score, and Accuracy are used as evaluation metrics for the query mapping module. 
 
If TP, TN, FP, and FN denote the true positives, true negatives, false positives, and false negatives respectively, then:

\begin{equation}
\label{eq:prec}
    Precision = \frac{TP}{TP + FP}
\end{equation}

\begin{equation}
\label{eq:rec}
    Recall = \frac{TP}{TP + FN}
\end{equation}

\begin{equation}
\label{eq:f1}
    F1-score =  \frac{2 \times Precision \times Recall}{Precision + Recall}
\end{equation}

As we have a multi-class classification problem, we use the weighted average of precision, recall, and F1-score for evaluation by computing the average of binary metrics weighted by the number of samples of each class as shown in Equation-\ref{eq:weighted}.
\begin{equation}
\label{eq:weighted}
  weighted\_average(X_l) = \frac{1}{\sum_{l \in L} n_l} \sum_{l \in L} n_l X_l
\end{equation}
Here $X_l$ denotes an evaluation metric for class $l$, $n_l$ denotes the number of samples for class $l$, $L$ is the set of classes, and $l$ is a single class such that $l \in L$.

Accuracy is calculated by dividing the number of samples that are correctly classified by the total number of samples in the dataset. This metric provides an overall measure of how well the model is able to predict the correct class for each instance.

\begin{equation}
    Accuracy = \frac{\textit{Number of correctly classified samples}}{\textit{Total number of samples}}
\end{equation}

Weighted recall and accuracy used for KisanQRS come out to be equal mathematically, which is also visible in the results. We choose to keep both in the results for clarity.

\item{\textit{Evaluation Metrics for Answer Retrieval and Ranking}}\\
We use the Normalized Discounted Cumulative Gain (NDCG) and Mean Average Precision (MAP) metrics to evaluate the performance of our retrieval system. Their brief description is given as follows:
\begin{enumerate}
    \item{\textit{NDCG}:  NDCG is a popular evaluation metric because it takes into account both the relevance of the items returned and their order in the ranking. This makes it a suitable metric for comparing the performance of different retrieval systems.

    \begin{equation}
        DCG = \sum_{i=1}^{n} \frac{2^{rel_i} - 1}{\log_2(i + 1)}
    \end{equation}
    \begin{equation}
        IDCG = \sum_{j=1}^{n} \frac{2^{rel_j} - 1}{\log_2(j + 1)}
        \end{equation}
    \begin{equation}
        NDCG = \frac{DCG}{IDCG}
    \end{equation}
    
    Here, $rel_i$ is the relevance score of the $i^{th}$ item in the ranking, $i$ is the position of the $i^{th}$ item in the ranking, starting from 1, $n$ is the total number of items in the ranking, $DCG$ is the Discounted Cumulative Gain, which is the sum of the discounted relevance scores of the items in the ranking, $rel_j$ is the relevance score of the $j^{th}$ item in the ideal ranking, $j$ is the position of the $j^{th}$ item in the ideal ranking, starting from 1, $IDCG$ is the Ideal Discounted Cumulative Gain, which is the maximum possible $DCG$ value for a given ranking.}

\item {\textit{MAP}:  
MAP is used in information retrieval to evaluate the performance of a model in retrieving relevant items from a large dataset. It provides an overall estimate of a model's accuracy at ranking a set of relevant items higher in the list of predicted items than the non-relevant items. The formula for MAP is given as follows:
    
    \begin{equation}
    \label{eq:map}
        \text{MAP} = \frac{1}{N} \sum_{i=1}^{N} \text{AP}_i
    \end{equation}
    
    where $N$ is the number of relevant items in the ground truth dataset, and $AP_i$ is the average precision of the $i^{th}$ relevant item. And the formula for average precision (AP) is:
    
    \begin{equation}
        \text{AP}_i = \frac{1}{\text{R}_i} \sum_{k=1}^{\text{R}_i} \text{P}(k) \cdot [\text{rel}(k) = 1]
    \end{equation}
    
    where, $R_i$ is the number of retrieved items for the $i^{th}$ relevant item, $P(k)$ is the precision at the $k^{th}$ retrieval, $rel(k)$ is the relevance of the item retrieved at the $k^{th}$ position, with $rel(k) = 1$ for relevant items and $rel(k) = 0$ for non-relevant items.}
\end{enumerate}   
\end{enumerate}

\subsubsection{Dataset}
\label{sec:subsa_dataset}
The dataset used for the evaluation of KisanQRS is collected from the KCC call logs repository. This repository contains records from 2006 to till date, totaling 34 million. A text record is kept of every query call made by farmers, including both, the information they request (query) and the information that is provided to them (response).     
For our experiments, we take data from QueryText, KccAns, and Crop attributes from 2018 to 2020 with the sample sizes mentioned in Table-\ref{table:table_dataset}.

\begin{table}[h!]
\centering
\captionsetup{justification=centering}
\caption{Sample Sizes used for Different Experiments}
\fontsize{10}{11}\selectfont
\begin{tabular}{l p{2.5cm} p{4cm} l}
\toprule
\textbf{Experiment} & \textbf{Sample Size} & \textbf{State(s)} & \textbf{Reference Results}   \\ \midrule
Evaluation of Query Clustering         & 12,000 & Andhra Pradesh & Table-\ref{table:table_cluster}  \\ \midrule
Model Selection for Query Mapping         & 12,000 & Andhra Pradesh & Table-\ref{table:table_deep}  \\ \midrule
Evaluation of Query Mapping   & 300,000 (60,000 per state) & Andhra Pradesh, Madhya Pradesh, Maharashtra, Tamil Nadu, Uttar Pradesh & Table-\ref{table:table_compare} \\ \midrule
Evaluation of Answer Retrieval         & 10,000  & Maharashtra, Tamil Nadu, Andhra Pradesh & Figure-\ref{fig:ndcg_chart}   \\  \bottomrule
\end{tabular}
\label{table:table_dataset}
\end{table}

\section{Results of the Study}
\label{sec:sa_ee}
In this section, we present the results of all three modules of KisanQRS: Clustering module, Query Mapping module, and Answer Retrieval module.

\subsection{Clustering Results}
\label{subsec:cluster_result}
In this subsection, we discuss the performance of different clustering algorithms on the basis of three evaluation metrics described in the previous section followed by a discussion about the runtime of these algorithms. Results are shown in Table-\ref{table:table_cluster}, where $\uparrow$ for a metric represents that a higher score is better, and $\downarrow$ denotes the lower, the better. The clustering algorithm of KisanQRS achieves a Silhouette score of 0.82 outperforming K-Means, Agglomerative, and DBSCAN, which represents that the query clusters are well-defined, with high intra-cluster similarity and low inter-cluster similarity. A high CH-index score of 4125.52 implies that the clusters are compact and well-separated, and a lower DB-Index score of 0.50 indicates less overlap and better-defined clusters.\\
The improved clustering performance of our method is due to the fact that it considers both semantic and token-wise similarity between queries to form corresponding clusters without any constraints on the total number of clusters.
K-Means partitions the dataset into a pre-defined number of clusters. It assumes that the data can be represented as spherical clusters with similar variances but it does not consider density or connectivity among data points. Hence, it struggles with clusters of varying sizes and shapes. As for DBSCAN, it is not suited for high-dimensional data as its clustering becomes less effective in higher-dimensional spaces. Agglomerative clustering is sensitive to the choice of distance metric and linkage criterion used to measure similarity between clusters. Different combinations of metrics and criteria lead to different clustering results. Overall the clustering method used for KisanQRS makes well-quality clusters that help to train the query-mapping module. 

\begin{table}[h!]
\centering
\captionsetup{justification=centering}
\caption{Performance Comparison of Clustering Techniques}
\begin{tabular}{l c c c c}
\toprule
\textbf{Clustering Technique} & \textbf{Silhouette Score($\uparrow$)} & \textbf{CH Index($\uparrow$)} & \textbf{DB Index($\downarrow$)} & \textbf{Runtime}   \\ \midrule
K-Means         & 0.7223  & 3528.7249   & 1.1212   &  19 min  \\ 
Agglomerative   & 0.7243  & 3450.6430    & 1.1348   & 14 min   \\ 
DBSCAN          & 0.6218  & 288.5214    & 0.9681   & 2 min   \\ 
KisanQRS      & \textbf{0.8228}  & \textbf{4125.5248}   & \textbf{0.4998}   & \textbf{12 sec}   \\ \bottomrule
\end{tabular}
\label{table:table_cluster}
\end{table}

The proposed algorithm works significantly faster than K-Means, DBSCAN, and Agglomerative clustering techniques, as represented in the time analysis graph in Figure-\ref{fig:time}. The proposed algorithm has a lower runtime because it takes advantage of the dataset's characteristic where queries for a specific topic mostly become the same after preprocessing. This reduces the need for extensive similarity comparisons, as queries that are similar to a common query are likely to be similar to each other as well.

\begin{figure}[h!]
    \centering
    \includegraphics[width = 10cm, height=7.5cm]{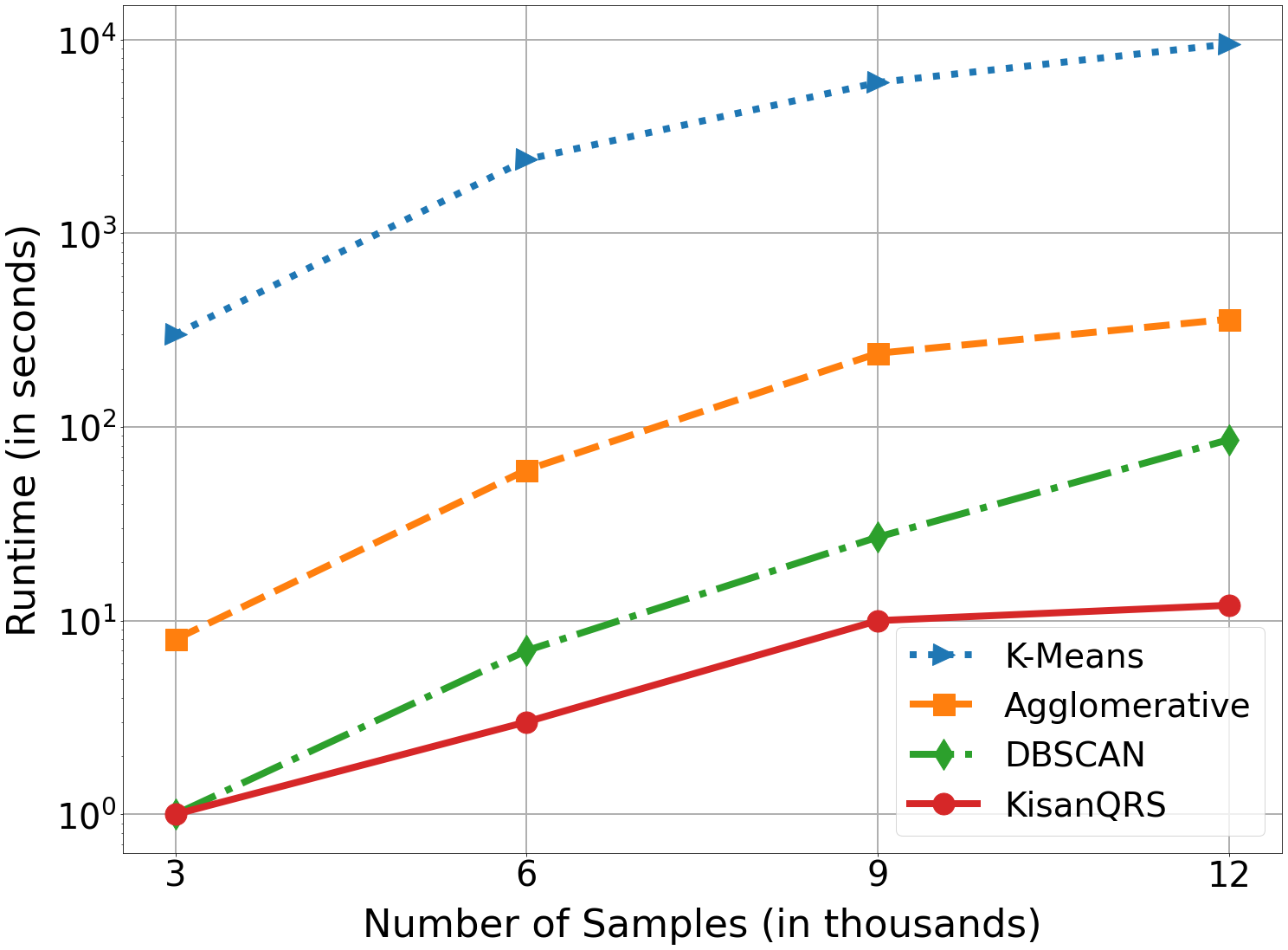}
    \caption{Runtime Comparison of Different Clustering Algorithms}
    \label{fig:time}
\end{figure}

\end{enumerate}

\subsection{Query Mapping Results}
\label{subsec:query_mapping results}

First, we conduct experiments to select the most suitable DL or ML model for the query mapping module. Subsequently, we evaluate the performance of our query mapping module by comparing its performance with other techniques. 

\begin{table}[h!]
\captionsetup{justification=centering}
\caption{Model Selection Results for Query Mapping for Different ML/DL Models}
\centering
\begin{tabular}{clc c c c}
\toprule
\textbf{Model} & \textbf{Input Type} & \textbf{Accuracy} & \textbf{Precision} & \textbf{
Recall} & \textbf{F1-score} \\ \midrule
{\multirow{5}{*}{MLP}} & SBERT & 96.61 & 95.86 & 96.61 & 96.02 \\
& Word2vec	&	96.63	&	96.12	&	96.63	&	96.13\\
& GloVe	&	95.43	&	95.06	&	95.43	&	94.96	\\
& BoW	&	96.20	&	95.61	&	96.20	&	95.67	\\
& TF-IDF	&	95.83	&	95.27	&	95.83	&	95.27	\\ \midrule

 {\multirow{5}{*}{LSTM}} & SBERT	&		\textbf{97.12}	&	96.34	&	\textbf{97.12}	&	\textbf{96.52}\\
&	Word2vec	&	96.68	&	\textbf{96.43}	&	96.68	&	96.3	\\
&	GloVe	&	95.62	&	95.05	&	95.62	&	95.13	\\
&	BoW	&	95.88	&	95.41	&	95.88	&	95.31	\\
&	TF-IDF	&	95.41	&	94.73	&	95.41	&	94.81	\\ \midrule

 {\multirow{5}{*}{GRU}} &	SBERT	&	96.58	&	96.13	&	96.58	&	96.08	\\
&	Word2vec	&	96.73	&	96.42	&	96.73	&	96.35	\\
&	GloVe	&	95.79	&	94.93	&	95.79	&	95.14	\\
&	BoW	&	95.92	&	95.34	&	95.92	&	95.36	\\
&	TF-IDF	&	95.42	&	94.87	&	95.42	&	94.83	\\ \midrule

 {\multirow{5}{*}{SVM}} &	SBERT	&	93.38	&	90.02	&	93.38	&	91.36	\\
&	Word2vec	&	95.47	&	93.81	&	95.47	&	94.36	\\
&	GloVe	&	93.69	&	91.23	&	93.69	&	92.13	\\
&	BoW	&	95.42	&	93.69	&	95.42	&	94.27	\\
&	TF-IDF	&	94.37	&	92.32	&	94.37	&	93.06	\\ \midrule

  {\multirow{5}{*}{LR}} &	SBERT	& 96.67	&	96.15	&	96.67	&	96.19	\\
&	Word2vec	&	96.03	&	95.10	&	96.03	&	95.29	\\
&	GloVe	&	95.43	&	95.06	&	95.43	&	94.96	\\
&	BoW	&	94.29	&	91.59	&	94.29	&	92.64	\\
&	TF-IDF	&	90.96	&	86.14	&	90.96	&	88.02	\\ \bottomrule

\end{tabular}
\label{table:table_deep}
\end{table}
\vspace{1mm}
\subsubsection{Model Selection Results for Query Mapping}
\label{subsubsec:model_selection}  
The experiments are conducted for five models out of which three are DL models: Multi-layer Perceptron (MLP), LSTM and Gated Recurrent Unit (GRU), and two are ML models: Support Vector Machine (SVM) and Logistic Regression (LR). Each of these models is tested for different input feature vectors, the set of which includes SBERT, Word2vec, GloVe, BoW, and TF-IDF. The performance of the models is assessed by employing accuracy, weighted precision, weighted recall, and weighted F1-score, which are widely used metrics for evaluating the performance of multi-class classification models. The results for these metrics are presented in Table-\ref{table:table_deep} and show that LSTM with SBERT is performing better as compared to other models. The best accuracy of 97.12\%, F1-score of 96.52\%, and recall of 97.12\% are achieved using the combination of LSTM and SBERT, while the highest recall is achieved using the combination of LSTM and Word2vec. \\
The superior performance of LSTM with SBERT can be attributed to several factors. First, SBERT is based on BERT architecture which is a transformer-based model. Transformer-based models consider the context of a word while extracting embeddings, whereas other feature extractors used in the comparison are context-independent methods. Secondly, DL models such as LSTM, GRU, and MLP excel at learning intricate patterns and representations from input feature vectors. Specifically, LSTM is adept at modeling sequential dependencies and capturing long-term contextual relationships. However, the dataset is also important for the selection of the appropriate method for the specific task. In the KCC dataset, many queries are in the form of short phrases where only the keywords convey the meaning of the query. For this reason, BoW, which is built around the unique words in the entire corpus, nearly matches the performance of SBERT. Yet, LSTM with SBERT provides slightly better performance than other combinations due to their aforementioned advantages. Hence, the LSTM with SBERT features is chosen for the query mapping module.

\begin{table}[h!]
\label{table:5states}
\captionsetup{justification=centering}
\caption{Evaluation of Query Mapping on 5 States Data}
\centering
\begin{tabular}{clc c c c}
\toprule

\textbf{State} & \textbf{Clustering Method} & \textbf{Accuracy} & \textbf{Precision} & \textbf{
Recall} & \textbf{F1-score} \\ \midrule

  {\multirow{6}{*}{Andhra Pradesh}} &	KisanQRS (0.80 thresh)	&	94.20	&	93.38	&	94.20	&	93.38	\\
&	KisanQRS (0.90 thresh)	&	96.35	&	\textbf{96.20}	&	96.35	&	96.00	\\
&	KisanQRS (0.95 thresh)	&	\textbf{96.64}	&	95.71	&	\textbf{96.64}	&	\textbf{96.01}	\\
&	K-Means	&	86.07	&	85.78	&	86.07	&	84.73	\\
&	Agglomerative	&	86.13	&	85.86	&	86.13	&	84.83	\\
&	DBSCAN	&	94.42	&	93.53	&	94.42	&	93.50	\\ \midrule

 {\multirow{6}{*}{Madhya Pradesh}} &	KisanQRS (0.80 thresh)	&	94.85	&	94.64	&	94.85	&	94.40	\\
&	KisanQRS (0.90 thresh)	&	96.41	&	96.03	&	96.41	&	95.94	\\
&	KisanQRS (0.95 thresh)	&	\textbf{97.03}	&	96.48	&	\textbf{97.03}	&	\textbf{96.58}	\\
&	K-Means	&	91.87	&	92.24	&	91.87	&	91.22	\\
&	Agglomerative	&	89.32	&	88.31	&	89.32	&	87.96	\\
&	DBSCAN	&	96.81	&	\textbf{96.60}	&	96.81	&	96.43	\\ \midrule

 {\multirow{6}{*}{Maharashtra}} &	KisanQRS (0.80 thresh)	&	96.08	&	95.56	&	96.08	&	95.58	\\
&	KisanQRS (0.90 thresh)	&	97.09	&	96.19	&	97.09	&	96.44	\\
&	KisanQRS (0.95 thresh)	&	\textbf{96.78}	&	95.57	&	\textbf{96.78}	&	\textbf{95.98}	\\
&	K-Means	&	89.70	&	88.83	&	89.70	&	88.65	\\
&	Agglomerative	&	88.96	&	87.63	&	88.96	&	87.45	\\
&	DBSCAN	&	95.82	&	\textbf{95.67}	&	95.82	&	95.37	\\ \midrule

{\multirow{6}{*}{Tamil Nadu}} &	KisanQRS (0.80 thresh)	&	92.07	&	91.30	&	92.07	&	90.99	\\
&	KisanQRS (0.90 thresh)	&	93.68	&	92.69	&	93.68	&	92.66	\\
&	KisanQRS (0.95 thresh)	&	\textbf{94.81}	&	\textbf{94.42}	&	\textbf{94.81}	&	\textbf{94.12}	\\
&	K-Means	&	85.77	&	87.01	&	85.77	&	85.08	\\
&	Agglomerative	&	87.87	&	87.65	&	87.87	&	86.88	\\
&	DBSCAN	&	93.80	&	93.83	&	93.80	&	93.30 \\
\midrule

 {\multirow{6}{*}{Uttar Pradesh}} &	KisanQRS (0.80 thresh)	&	88.27	&	88.91	&	88.27	&	87.63	\\
&	KisanQRS (0.90 thresh)	&	93.49	&	93.33	&	93.49	&	92.86	\\
&	KisanQRS (0.95 thresh)	&	\textbf{94.59}	&	\textbf{93.77}	&	\textbf{94.59}	&	\textbf{93.86}	\\
&	K-Means	&	74.51	&	76.45	&	74.51	&	73.62	\\
&	Agglomerative	&	73.65	&	73.49	&	73.65	&	71.79	\\
&	DBSCAN	&	93.67	&	94.38	&	93.67	&	93.41	\\ \bottomrule

\end{tabular}
\label{table:table_compare}
\end{table}

\vspace{1mm}
\subsubsection{Evaluation Results of Query Mapping Module}
\label{subsubsec:eval_query}
The performance of the query mapping module of KisanQRS is evaluated on five different states' data, namely Andhra Pradesh, Madhya Pradesh, Maharashtra, Tamil Nadu, and Uttar Pradesh. Since the performance of the query mapping module is affected by the quality of clusters formed 
as the cluster labels eventually serve as the target labels for the LSTM-based query mapping module, the scope of our experiments includes K-Means, Agglomerative, DBSCAN, and proposed KisanQRS clustering algorithm for three $thresh$ values; 0.80, 0,90 and 0.95. A higher value of $thresh$, like $0.99$, produces a large number of clusters, with each cluster containing identical queries, which eliminates the queries having the same semantic meaning but different lexical structures to be placed in the same cluster. In contrast, a lower $thresh$ value may result in clusters containing queries that are not semantically similar. Therefore, a moderate $thresh$ value such as 0.95 is the optimal choice for KisanQRS. The results are shown in Table-\ref{table:table_compare}. KisanQRS query mapping outperforms K-Means and Agglomerative clustering algorithms by a high margin, whereas DBSCAN is close to its performance but still slightly lower. These results indicate that KisanQRS accurately maps a new query to the correct cluster, which ultimately helps in retrieving the precise responses for the given query.\\

\begin{figure}[htbp]
  \centering
  \subfigure[Andhra Pradesh]{\includegraphics[width=0.47\textwidth]{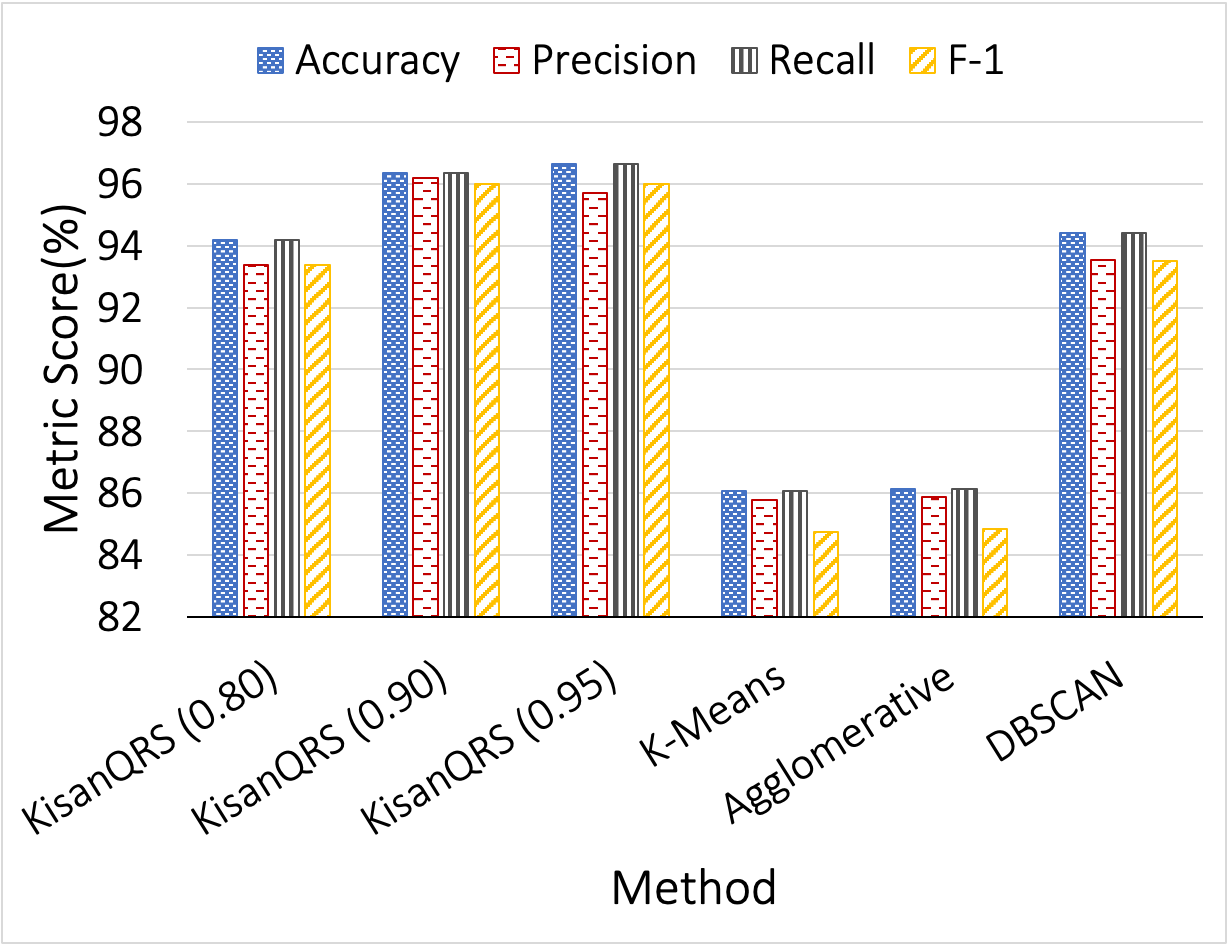}}
  \hfill
  \subfigure[Madhya Pradesh]{\includegraphics[width=0.47\textwidth]{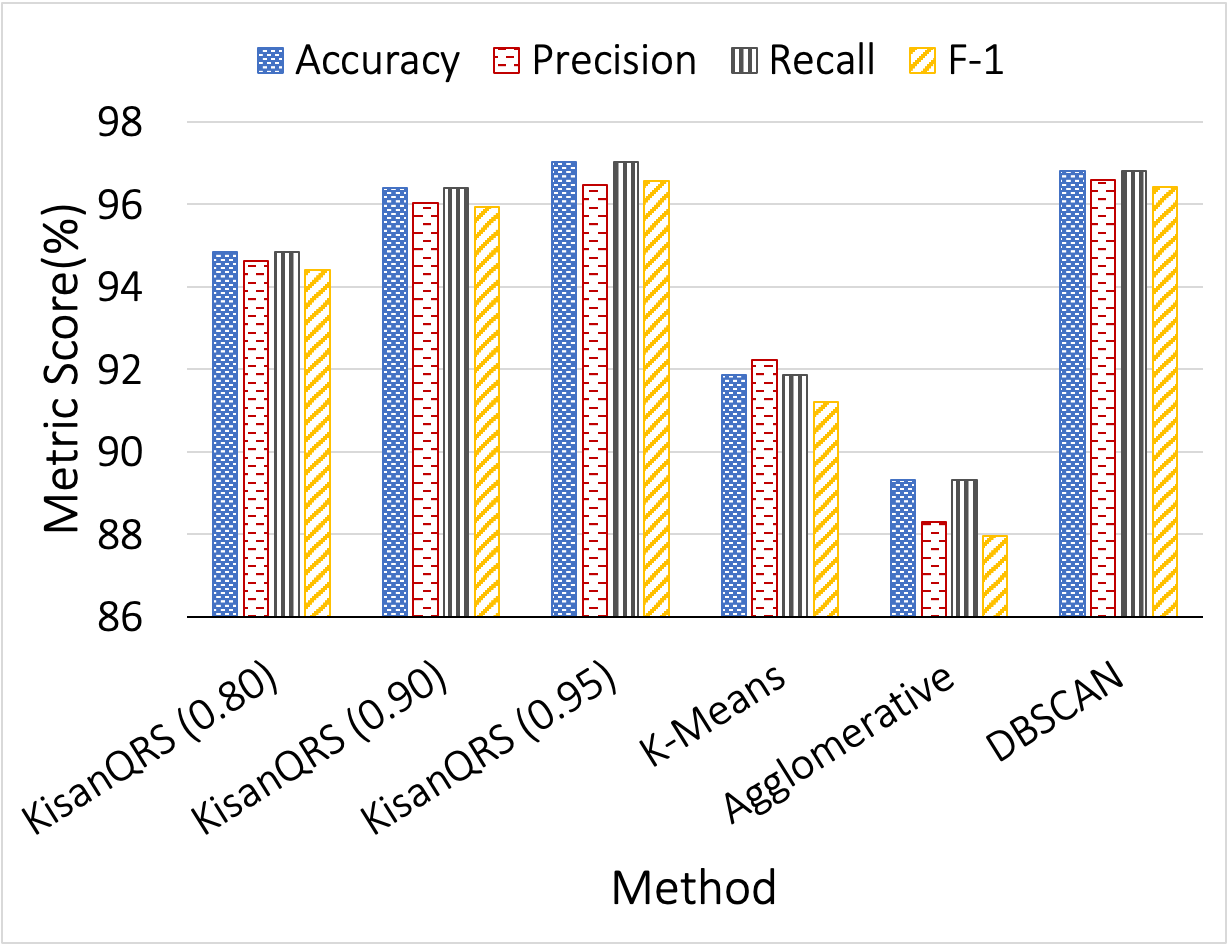}}
  \hfill
  \subfigure[Maharashtra]{\includegraphics[width=0.47\textwidth]{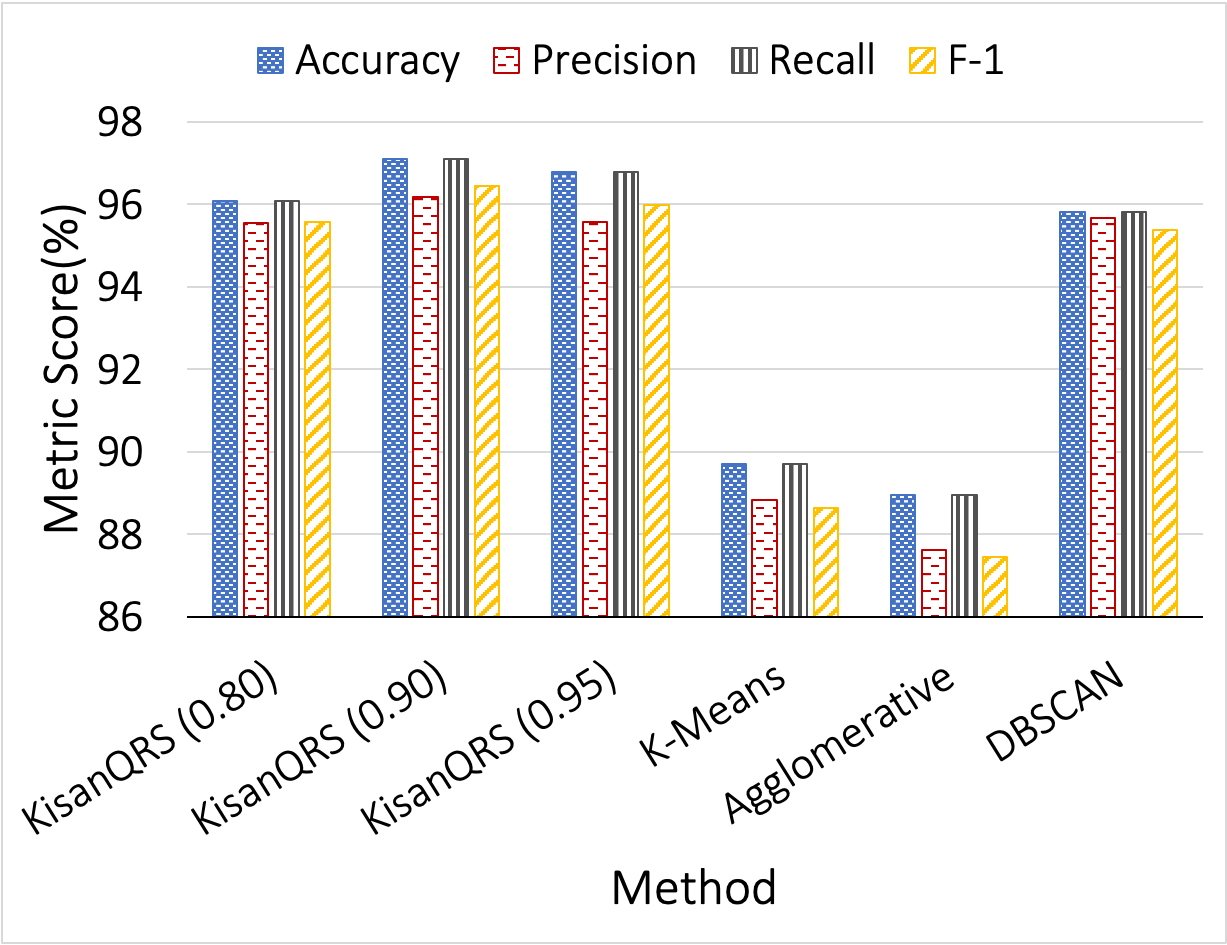}}
  \hfill
  \subfigure[Tamil Nadu]{\includegraphics[width=0.47\textwidth]{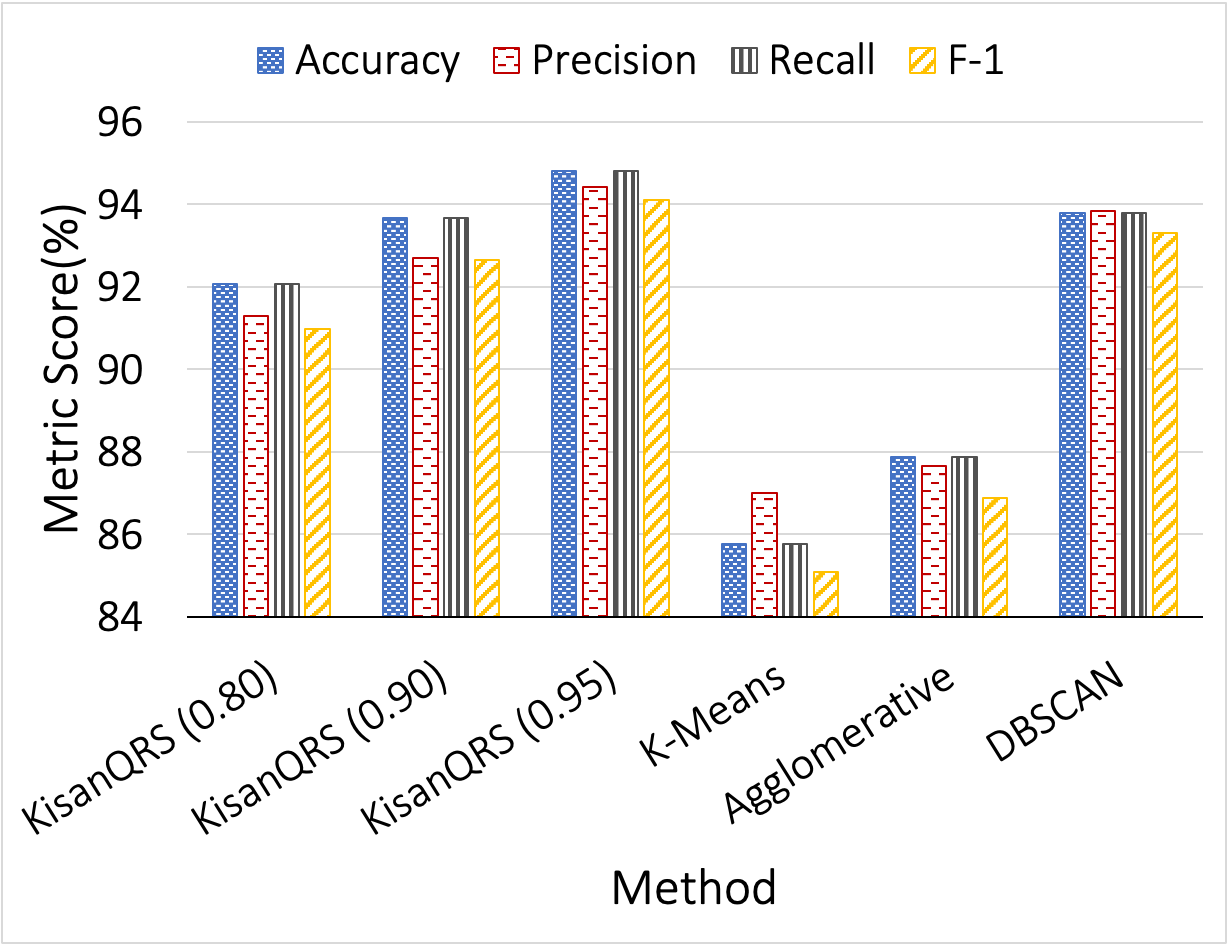}}
  \hfill
  \subfigure[Uttar Pradesh]{\includegraphics[width=0.47\textwidth]{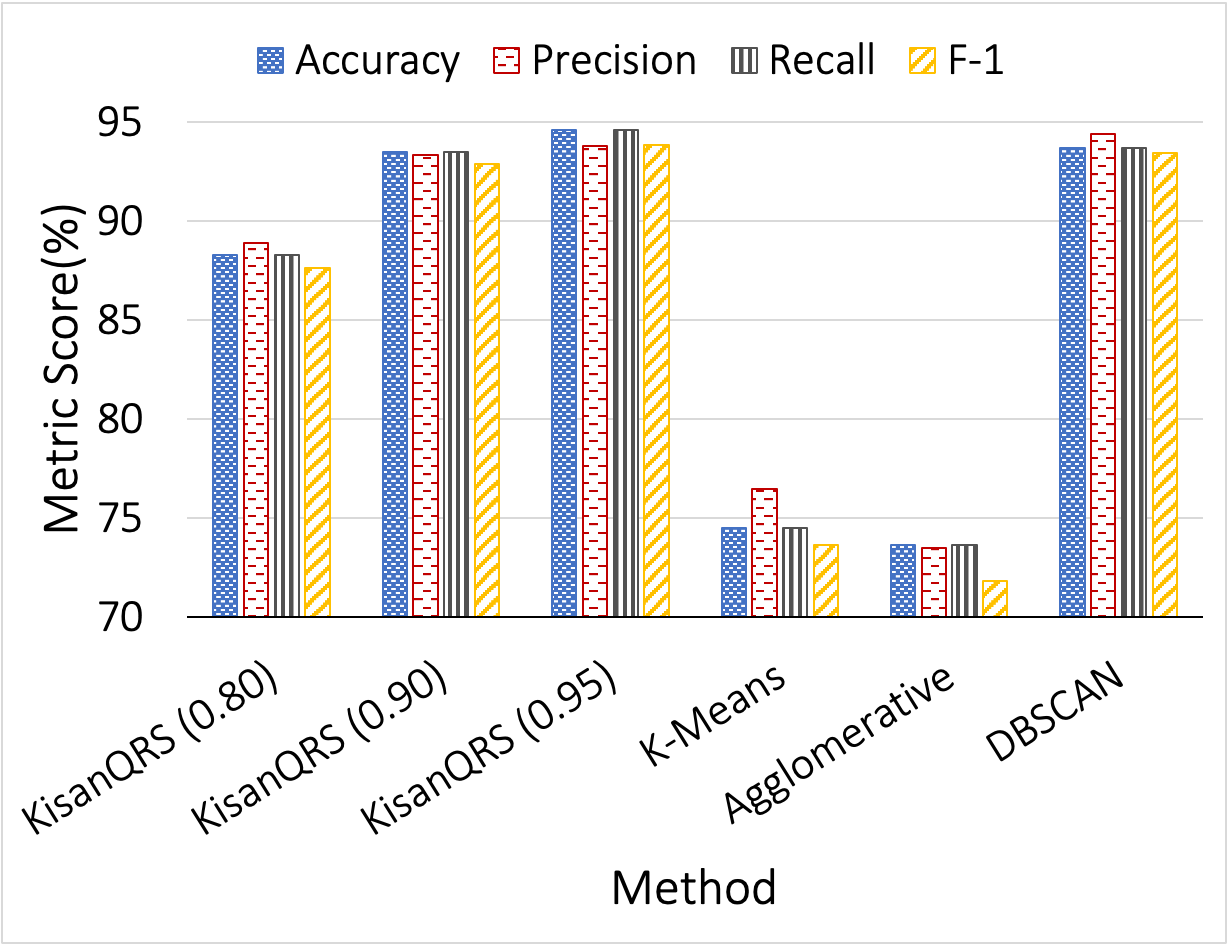}}
  \caption{Performance Comparison of KisanQRS Query Mapping on 5 States Data}
  \label{fig:5state_chart}
\end{figure}

The major factor for KisanQRS query mapping outperforming other methods is that query mapping depends on the quality of clusters. The KisanQRS query mapping uses cluster labels using its own clustering method, which provides better clusters than the other methods used for comparison.
A graphical representation of these results is given in Figure-\ref{fig:5state_chart}.

\subsection{Answer Retrieval Results}
\begin{figure}[h!]
    \centering
    \includegraphics[width = 10cm, height=7.5cm]{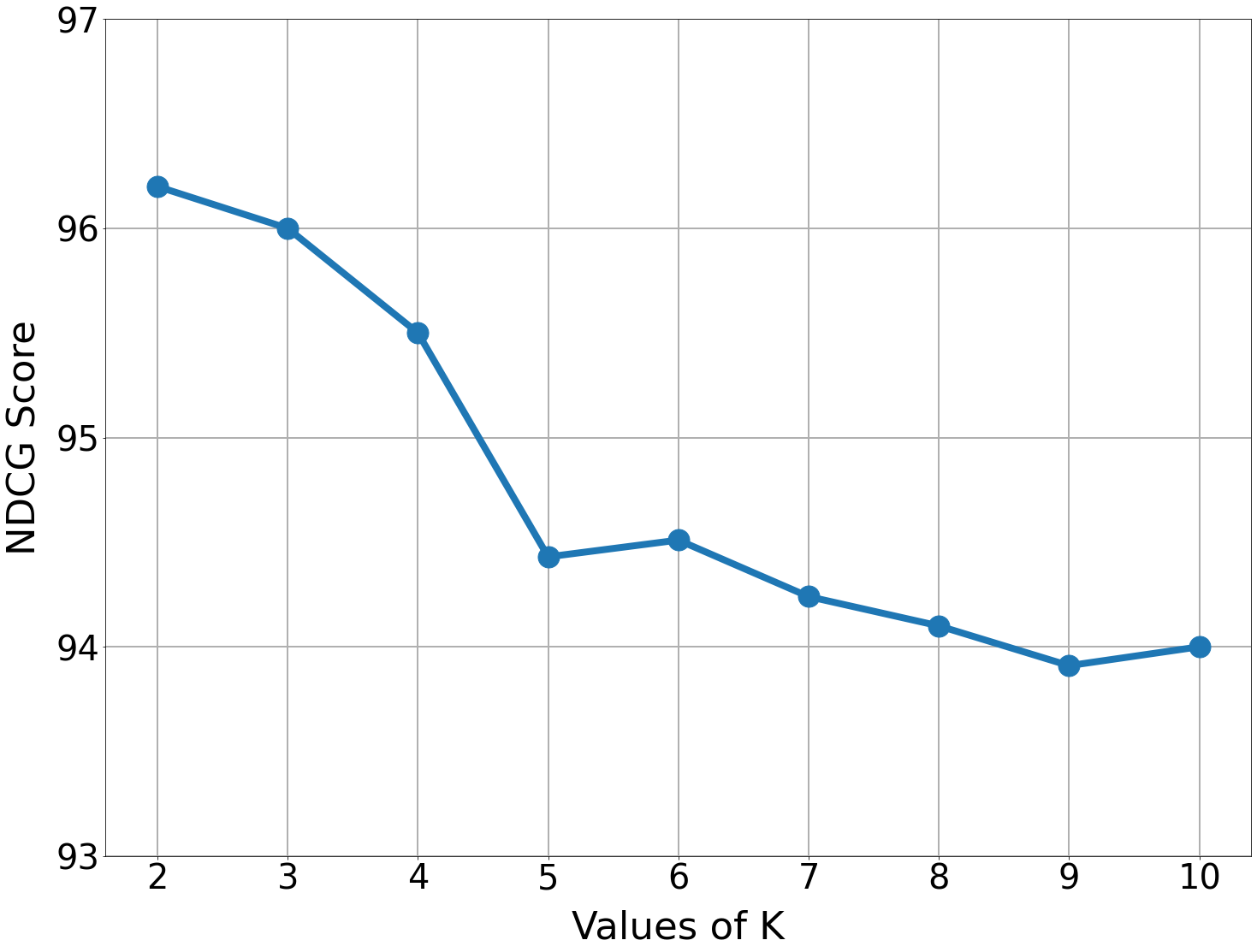}
    \caption{NDCG Scores for Answer Retrieval on Different Values of K}
    \label{fig:ndcg_chart}
\end{figure}

Results for answer retrieval are taken on the set of answers which are scored on a scale of 0-10, i.e. $rel_i \in \{0, 1, 2, 3, 4, 5,6,7,8,9,10\}$, where 10 denotes `highly relevant' and 0 denotes `not relevant' to the corresponding query. For different values of $K$, the answers are re-ranked in the range of $0$-$K$ on the basis of their relative ranking in the originally scored set. For each query, a set of top-$K$ answers is generated by KisanQRS, and NDCG scores are computed for the KisanQRS generated ranking of answers, where re-ranked answers for the same value of $K$ serve as the ground truth. 
The NDCG scores trend for different values of $K$ is shown in Figure-\ref{fig:ndcg_chart}.

For MAP score, If a query, $q_{user}$, is mapped to the correct query cluster, all the retrieved answers by the retrieval system will be relevant, i.e. $AP_i = 1$ (Equation-\ref{eq:map}). This is due to the fact that each query cluster has semantically and lexically similar queries, hence each answer of that cluster is the potential response. However, If a query is not mapped to the correct query cluster, all the answers retrieved will be irrelevant, i.e., $AP_i = 0$ (Equation-\ref{eq:map}) in that case. This indicates that the MAP is equivalent to the weighted precision of the LSTM-based query mapping module, which is $96.34\%$. 
The model is capable of generating top-$K$ answers, given $q_{user}$. The top-5 answers retrieved by the model for two user queries, "What is the fertilizer dose for mosambi?" and "What to do if pink bollworm attacks cotton?" are shown in Table-\ref{table:table_query1} and \ref{table:table_query2}.

\begin{table}[h!]
\centering
\captionsetup{justification=centering}
\caption{Top-5 Retrieved Answers for the Query "What is the fertilizer dose for mosambi?"}
\fontsize{10}{11}\selectfont
\begin{tabular}{cllp{8cm}}
\toprule
\textbf{Rank} & \textbf{Crop} & \textbf{QueryText} & \textbf{KccAns} \\ \midrule
1 & Mosambi & Fertilizer dose for Mosambi & Fertilizer dose for Mosambi- 151515 400 gm neem pend 1 kg cow dung 20 kg plant \\ \midrule
2 & Mosambi & Fertilizer dose for Mosambi & Spray 00:52:34 60 gm 15 liter water \\ \midrule
3 & Mosambi & Fertilizer dose for Mosambi & Fertilizer dose for Mosambi : - 5 kg fym 120 kg nitrogen 60 kg phosphorus 30 gm micronutrient plant \\ \midrule
4 & Mosambi & Fertilizer dose for Mosambi & Fertilizer dose for Mosambi : 100 gram urea 100 gram ssp 500 gram 500 gram nimoli pend per tree \\ \midrule
5 & Mosambi & Fertilizer dose for Mosambi & Fertilizer dose for Mosambi : - 10:26:26 50 kg 20:20:00:13 50 kg micronutrient 7 kg \\ \bottomrule
\end{tabular}
\label{table:table_query1}
\end{table}

\begin{table}[h!]
\centering
\captionsetup{justification=centering}
\caption{Top-5 Retrieved Answers for the Query "What to do if pink bollworm attacks cotton kapas?"}
\fontsize{10}{11}\selectfont
\begin{tabular}{cllp{6.3cm}}
\toprule
\textbf{Rank} & \textbf{Crop} & \textbf{QueryText} & \textbf{KccAns} \\ \midrule
1& Cotton Kapas & Control of Pink Bollworm of Cotton & Use pheromones and light traps to set one or more 10 bird stops. Sprinkle neem oil at the beginning and sprinkle profhenophos 2 ml per liter of water. Collect occasionally pestered leaves, flowers, and bolls and destroy them. \\ \midrule
2 & Cotton Kapas & Attack of Pink Bollworm on Cotton & Spray profex super 30 ml/15 liter water and profenofos 40 cypermethrin/4 Nagarjuna. \\ \midrule
3 & Cotton Kapas & Attack of Pink Bollworm on Cotton & Spray proclaim 5 gm/15 lit of water and emamectin benzoate 5 sg-Syngenta. \\ \midrule
4 & Cotton Kapas & Attack of Pink Bollworm on Cotton & Spray neem ark 30 ml/15 liter water. \\ \midrule
5 & Cotton Kapas & Attack of Pink Bollworm on Cotton & Spray quinolphos 30 ml and bavistin 30 gm/15 liter water. \\ \bottomrule
\end{tabular}
\label{table:table_query2}
\end{table}

\section{Discussion}
\label{sec:sa_disc}
This article proposes a novel approach to retrieve top-$K$ answers for a farmer’s query. The relevance of retrieved answers depends directly on the performance of the DL model to correctly map the user query to its corresponding cluster and indirectly on the quality of clusters formed as they serve as the labels for training the DL model. We combine the cosine similarity scores of embeddings with the Jaccard similarity scores to form the similarity matrix, which acts as a basis for clustering. The used proportion of these scores is contingent upon the states as the manner in which queries are formulated for different states is different. The threshold parameter of clustering is also experimented with multiple values, and it is observed that when the LSTM-based query mapping is trained on cluster labels generated with a threshold value greater than 0.8, it performs better as compared to lower threshold values. Our proposed clustering algorithm achieves a Silhouette Score of 0.82, which suggests that the clusters are well-separated and distinct from each other. Furthermore, the achieved accuracy of 97.12\% and F1-score of 96.5\% for the LSTM model indicates that the model is able to effectively learn and generalize from the training data to make accurate predictions on the test data. The performance of the LSTM-based network can be attributed to the fact that LSTM is specifically designed to handle sequential data by capturing long-term dependencies. It can effectively model the contextual relationships between words in a sentence, allowing it to capture better the semantic structure of the text, which gives it an edge over other DL methods in text classification tasks. ML methods often treat words as independent features and fail to model the sequential or hierarchical structure of the text, resulting in a limited ability to understand the true meaning of the text. However, it is worth noting that the selection of DL and traditional ML approaches also depends on specific task characteristics and datasets. In our study, we observe that LSTM, GRU, and MLP have similar performance in query mapping, with LSTM performing slightly better with approximately 0.5\% better accuracy. Considering these results and the aforementioned advantages, we select the LSTM network for the query mapping module.

\subsection{Limitations}
The KisanQRS has a few limitations that can be addressed in future works. Some of the queries and answers do not have a clear semantic meaning and mostly keywords are there. Hence, from the perspective of the DL network, it becomes a bottleneck for the DL method if sentences do not have clear semantic meaning.  In such sentences, the LSTM-based network struggles to model the contextual relationships between words. To overcome this challenge, the queries and answers need to be formulated such that some meaning can be inferred. Moreover, scientific expertise is required to identify the most suitable answers to address the user query in the given situation. Additionally, KisanQRS is not capable of providing answers to queries that require real-time information, such as market rates of crops and weather conditions. Answers to such queries change very frequently since they depend on real-time information. Ensuring the model's robustness against such errors requires further efforts beyond the scope of our current work.

\subsection{Practical Implications}
 A platform based on our model that addresses daily agricultural queries with a reliable, data-driven response system would greatly benefit farmers, leading to better decision-making and awareness regarding the best practices. The platform could also be integrated into a voice assisted chatbot, allowing farmers who may be digitally under-skilled to simply speak their queries and receive answers in audio form. Call centre workers can also utilize this system to access information swiftly and accurately to address inquiries.

\section{Conclusion}
\label{sec:sa_cnfw}
The inadequate access to technology in numerous developing nations has resulted in the absence of a sturdy system for examining the challenges encountered by farmers. KCCs in India have played a critical role in disseminating information to farmers through telephonic responses, but the lack of expertise and availability of call centre agents limits the usefulness of farmer's helpline centres. In this direction, we introduce KisanQRS, a novel Deep Learning-based approach to provide quick and pertinent responses to farmers' queries. KisanQRS pipeline employs a robust clustering technique, a Deep Learning-based query mapping module, and a novel answer retrieval module. The clustering technique employed in KisanQRS delivers significantly faster performance and surpasses K-Means, Agglomerative, and DBSCAN in terms of cluster quality. An extensive evaluation of various DL and ML models, such as MLP, LSTM, GRU, SVM, and LR, is conducted to identify the best-suited method for query mapping. Experimental results indicate that the LSTM model outperforms other models in terms of performance by achieving the highest accuracy
when trained on cluster labels generated using the proposed clustering method. Moreover, the answer retrieval technique achieves high NDCG score in retrieving top-$K$ answers. The dataset for experiments is taken from the call logs of KCC, a farmer's helpline centre in India. Overall, the experimental results demonstrate the efficacy of KisanQRS in providing quick and relevant responses to agricultural queries. Adopting KisanQRS can help automate the query-response system and enhance the farmers' agricultural decision-making. 

\bibliographystyle{cas-model2-names}

\bibliography{sample}

\end{document}